\documentclass[11pt,dvips,a4]{article}
\usepackage{amssymb}
\usepackage{amsmath}
\usepackage{color}
\usepackage[colorlinks]{hyperref}

\newcommand{\bra}{\begin{array}}
\newcommand{\era}{\end{array}}
\newcommand{\beq}{\begin{equation}}
\newcommand{\eeq}{\end{equation}}
\newcommand{\beqar}{\begin{eqnarray}}
\newcommand{\eeqar}{\end{eqnarray}}

\def\BC{\bb C}
\def\_\BC{\bbi C}

\def\Tr {{\rm Tr}}

\def\( {\left(}
   \def\) {\right)}
\def\[ {\left[}
\def\] {\right]}
\def\no2 {{\textstyle{n\over 2}}}

\def\Tr {{\rm Tr}}
\def\dag {{\dagger}}
\newcommand{\om}{\omega}
\newcommand{\Om}{\Omega}
\newcommand{\lam}{\lambda}

\newcommand{\ot}{\otimes}

\newcommand{\be}{\beta}

\newcommand{\ga}{\gamma}
\newcommand{\te}{\theta}

\newcommand{\pa}{\partial}
\newcommand{\al}{\alpha}

\newcommand{\lga}{\longrightarrow}

\newcommand{\ka}{\kappa}

\newcommand{\ti}{\tilde}
\newcommand{\da}{\dagger}

\newcommand{\ov}{\over}
\newcommand{\hb}{\hbar}

\newcommand{\sq}{\sqrt}

\newcommand{\lb}{\label}

\begin{document}
\begin{titlepage}
\setcounter{page}{1}
\renewcommand{\thefootnote}{\fnsymbol{footnote}}

\begin{flushright}
ucd-tpg:09.04~~~~~\\
arXiv:0911.2837
\end{flushright}

\vspace{5mm}
\begin{center}

%\begin{document}
%\thispagestyle{empty}

%\begin{flushright}
%ucd-tpg/06xx\\
%hep-th/yymmxxx
%\end{flushright}

\vspace{0.5cm}
%\begin{center}
 {\Large \bf Magnetism of Two Coupled Harmonic Oscillators}\\ % with Magnetic Field}\\

\vspace{0.5cm}

{\bf Mohammed Daoud$^{a}$, Mohamed El Bouziani$^{b}$},
{\bf Rachid Hou\c{c}a$^{b}$}%\footnote{rachid.houca@gmail.com}}
and {\bf Ahmed Jellal$^{b,c,d}$\footnote{{\sf jellal@pks.mpg.de} and {\sf jellal@ucd.ma}}}\\
\vspace{0.5cm}

{ \em $^{a}$Physics Department, Faculty of Sciences, University Ibn Zohr,\\ 
PO Box 8106, Agadir,
Morocco}

{ \em $^{b}$Theoretical Physics Group,  %Department of Physics,
Faculty of Sciences, Choua\"ib Doukkali University,\\
%{\em Ibn Ma\^achou Road, 
PO Box 20, 24000 El Jadida,
Morocco} %\\[1em] \]

{\em $^{c}${\it Physics Department, College of Sciences, King Faisal University,\\   
PO Box 9149, Alahssa 31982,
Saudi Arabia}}

{\em$^{d}$Max Planck Institute for the Physics of Complex Systems,\\
N\"othnitzer Str. 38, 01187 Dresden, Germany} 
\\[1em]

\vspace{3cm}

\begin{abstract}

%We analyze
The thermodynamical properties of a system
of two coupled harmonic oscillators in the presence of
an uniform magnetic field $B$ are investigated. Using an unitary transformation,
we show that the system can be diagonalized in simple way
and then obtain the energy spectrum solutions. % in terms of the coupling parameter $\al$.
These will be used to determine the thermodynamical potential
in terms of different physical parameters like the coupling
parameter $\al$.
This allows us to give a generalization
of already significant published work and obtain different results,
those could be used to discuss the magnetism of the system. Different limiting
cases, in terms of $\al$ and $B$, have been discussed. In fact,
 quantum corrections to the Landau diamagntesim and orbital paramagnetism
are found.
%This has been done by evaluating different thermodynamics properties
%of the present system.

\end{abstract}
\end{center}
\end{titlepage}

\newpage
%%%%%%%%%%%%%%%%%%%%%%%%%%%%%
\section{Introduction}
%%%%%%%%%%%%%%%%%%%%%%%%%%%%%%

Since the pioneering work of Landau in 1930~\cite{landau}, orbital
magnetism of electron gases has been the subject of considerable
attention, especially during the last decades with the
advent of experimental opportunities, more precisely with
the availability of two-dimensional electronic devices, quantum
boxes, or mesoscopic finite-size objects. One can find in
\cite{ber} or \cite{fks}  a good account of the theoretical investigations
on the subject, especially from a semiclassical point of
view. Fore a more recent developments, we cite the book~\cite{Plischke}

%Specifically,
Fukuyama group has been developed an amount of papers
dealing with different features of two-dimensional systems.
Among them, we cite the reference~\cite{yfu} where
%For instance,
%they investigated the
the magnetization of such systems in external potentials~\cite{yfu}
are studied. In fact, the
magnetic field $B$ and temperature $T$ dependence of the magnetization is calculated exactly.
 It is
found that the magnetization is well defined in the limit of vanishing $B$ as well
as in the limit of $T=0$, which showing a large fluctuation at low $T$ as $B$
 is varied. It is shown that this fluctuating magnetization tends to the Landau diamagnetism
at higher $T$ or by ensemble averaging.
Subsequently, other exciting investigation has been reported on~\cite{ifu}, which
concerned
the spatial distribution of electric current under $B$ and the resultant orbital
magnetism for the present system under a harmonic confining potential
$V(\vec{r})=m \omega_0^2 r^2/2$ in various regimes of the couple $(T,B)$. As an interesting
result is that the
microscopic conditions for the validity of Landau diamagnetism are clarified. %~\cite{ifu}.
% Under a weak magnetic
%field $(\omega_c\leq\omega_0, \omega_c$ being a cyclotron frequency) and at low temperature
%$(T\leq\hbar\omega_0)$, where the orbital magnetic moment fluctuates as a function of the field,
%the currents are irregularly distributed paramagnetically or diamagnetically inside the bulk region.
%As the temperature is raised under such a weak field, however, the currents in the bulk region are
%immediately reduced and finally there only remains the diamagnetic current flowing along the edge.
%At the same time, the usual Landau diamagnetism results for the total magnetic moment. The origin of
%this dramatic temperature dependence is seen to be in the multiple reflection of electron waves by
%the boundary confining potential, which becomes important once the coherence length of electrons gets
%longer than the system length. Under a stronger field $(\omega_c\geq\omega_0)$, on the other hand,
%the currents in the bulk region cause de Haas-van Alphen effect at low temperature as $T\leq\hbar\omega_c$.
%As the temperature gets higher $(T\geq\hbar\omega_c)$ under such a strong field, the bulk currents are reduced
%and the Landau diamagnetism by the edge current is recovered.

The system studied in~\cite{ifu} has been considered from another point of view.
More precisely, a coherent states approach is used to investigate
its basic features~\cite{Gazeau}. In fact,
%The above systems analyzed again in terms of the coherent states approach. Indeed,
the corresponding expressions for the thermodynamical
potential and magnetic moment are determined. These are exact, in contrast to those in reference~\cite{ifu},
and the results yielded a full description of the phase diagram
of the magnetization. The derivation crucially rests
upon the observation that the Fermi-Dirac function is a fixed
point of the Fourier transform. Exact series expansions ensue
by simple application of the residue theorem. The related
physical quantities are obtained and different discussions are
reported in terms of the natures of $T$ and $B$.
%in different regime of temperature and magnetic field.
These concern the thermodynamical potential, the orbital magnetic moment,
the subsequent magnetic susceptibility and the average number
of electrons.
% are
%easily tractable and analyzable.

On the other hand, the problem of two coupled harmonic oscillators
living on two dimensions  was investigated at different
occasions where several papers are developed by Kim group,
 for a short list we cite~\cite{kim}-\cite{kimg}.
Furthermore, the quantum mechanical of such systems
on the non-commutative plane has been studied as well~\cite{jellal}
where different quantum corrections to the original work~\cite{kim}
are obtained and their interpretations are given.
%This has been done by demanding that the spatial
%coordinates do not commute. The star-product was used to write the non-commutative Hamiltonian and solve
%the eigenequations to get the eigenstates as well as the energy spectrum. Focusing on the ground
%state, the corresponding density matrix was evaluated. It depends on the non-commutativity
%parameter $\te$ but its traces as well as the entropy are not affected by the non-commutativity. Also
 %the Wigner function is calculated to confirm the $\te$-independence of the traces. The uncertainty
%relation on $R^2_\te$ is found to depend on θ and it coincides with the normal case in the limit $\te = 0$.
%For small $\te$, it is showen that this relation contains a quantum correction, which is a shift with a
%$\te^2$ term. Two limits of the mixing angle, namely $\frac{\pi}{ 2}$
%and $0$, have been discussed. In the last case, the system can be linked to the Hall electron.

After mentioning the above results, an interesting question
arises immediately that concerns other features of
two coupled harmonic oscillators. Specifically, it is possible to
study the thermodynamical properties of such systems in the presence
of an uniform magnetic field. The answer
will be the subject of the present paper where interesting results will
be derived and discussed. In fact, we will show how to use
the machinery developed by one of the present authors in the basic
reference~\cite{Gazeau} to analyze the magnetization of the  system.

%In the present work, we study another features of
%two coupled harmonic oscillators, which concerns its magnetism.
More precisely,
we develop
a theory that analyzes the basic features
of two coupled harmonic oscillators under the magnetic field. In doing so,
we inspect two already published works~\cite{Gazeau,jellal}
%of one of the present author
 to generate
a full description of the present system from
thermodynamical point of view. Actually, this can be done
by the help of the energy spectrum solutions. To derive them, we
%can be obatined
%getting the spectrum energy and the corresponding
%wavefunctions. These can be obtained
make use
of an unitary transformation that leads to a solvable
Hamiltonian of the system.

Subsequently, we present two ways to evaluate the
thermodynamical potential. Indeed, from the Berezin--Lieb inequalities and
after determining some physical quantities,
we discuss different limiting case in terms of the involved physical parameters.
%involved in the game and in particular, the coupling parameter.
These lead to end up with interesting results and in particular
we show that the average number of electrons behaves like
in the inverse of squared magnetic field for the
infinite coupling limit, i.e. $\al\lga\infty$. In this situation, the system behaves
like a quantum Hall effect one~\cite{prange}. On the other hand,
quantum corrections to the orbital paramagnetism and Landau diamagnetism
are obtained. More importantly, we notice that
by switching off $\al$ in our analysis, we recover
already published work~\cite{Gazeau}.

Furthermore, we give an exact formula
of the thermodynamical potential with the help of some well-known relations
and discuss different issues. Indeed, applying the Fermi--Dirac trace formulas,
we explicitly derive the average number of electrons
and the magnetic moment. Finally, we how that they can be reduced
to the standard expressions by taking into account the
the liming case %where their limiting cases are obatined for
$\al=0$.
%These concern
%the high and low temperature limits as well as
%the nature of the magnetic field.

The present paper is organized as follows. In section $2$,
we formulate our problem by establishing the necessary
materials to deal with our task. In section $3$, after
making use of an unitary transformation,
we introduce an algebraic method to derive the energy
spectrum solutions. Their underlying  properties
will be discussed by
considering four limiting cases.
%eigenvalues as well as the corresponding wavefunctions
%and we discuss different cases: week and strong magnetic
%field limits of the energy spectrum.
We construct
the coherent states for the present solutions
and show that they are coupling parameter dependent
in section $4$.
These will serve as tools to determine explicitly
different physical quantities and in particular
the thermodynamical potential in section $5$.
It will be obtained by adopting the Berezin--Lieb
inequalities in the first stage. This allows us to
give different discussion and end up with interesting
conclusions. However, in section $6$, we consider another approach based
on the Fermi--Dirac trace formulas
to give the exact form of the
 thermodynamical potential.
%and present different
%discussions in terms of the high and low temperature
%regimes.
Finally, we conclude and give different
perspectives. % in the last section.

%%%%%%%%%%%%%%%%%%%%%%%%%%%%%%%%%%%%%%%%%
\section{Formulating the problem}
%%%%%%%%%%%%%%%%%%%%%%%%%%%%%%%%%%%%%%%%%

We start by formulating our problem by setting the
needed tools for doing our
task. This can be done by establishing a mathematical formalism
governed by a Hamiltonian describing a system of two coupled harmonic
oscillators in two dimensions. Subsequently, we submit the
system to a constant magnetic field and analyze its behavior.
In doing so, we determine the
energy spectrum through an algebraic method after making use of
an unitary transformation.

%As we claimed before, we are wondering to
%investigate the basic features of two coupled harmonic oscillators where
%a minimal coupling is introduced. For this,
%we formulate our problem and use an
%unitary transformation to end up with
%a diagonalized Hamiltonian and of
%course derive the eigenvalues
%and eigenstates.

%%%%%%%%%%%%%%%%%%%%%%%%%%%%%%%%%%%%%%%%%%
\subsection{Coupled harmonic oscillators}
%%%%%%%%%%%%%%%%%%%%%%%%%%%%%%%%%%%%%%%%%%%%%%

We consider a system of two coupled harmonic oscillators of mass
$(m_1,m_2)$ and living on the plane $(X_1, X_2)$. This can be described by
a
Hamiltonian as sum of free
and interacting parts, such as
\beq\label{ham0}
H=\frac{P_1^2}{2m_1}+\frac{P_2^2}{2m_2}+\frac{1}{2}\left(C_1X_1^2+C_2X_2^2+C_3X_1X_2\right)
\eeq
where $C_1$, $C_2$ and $C_3$ are three constant parameters. Note in passing that,
the involved parameter can be fixed according to the nature of the
system. On the other hand, (\ref{ham0})
has been investigated for different purposes, for instance
one may see reference~\cite{kim}, which has been generalized to the non-commutative
geometry case~\cite{jellal}.

As claimed before, we are wondering to study the magnetization
of two coupled harmonic oscillators under an uniform magnetic field. To achieve this goal,
we generalize the system governed by (\ref{ham0})
to another one of Hamiltonian
\beq\label{ham1}
H_1={\Pi_1^2\over2m_1} +{\Pi_2^2\over2m_2}+{1\over2}\left(C_1X_1^2+C_2X_2^2+C_3X_1X_2\right)
\eeq
where $\Pi_1$ and $\Pi_2$ are the conjugate momentum.
They can be simplified by choosing an appropriate gauge. Indeed,
in the symmetric gauge
\beq
\vec{A}={B\ov2}\left(-X_2,X_1\right)
\eeq
they are given by
\beq
\Pi_1=P_1-{eB\over2c}X_2, \qquad \Pi_2=P_2-{eB\over2c}X_1.
\eeq
Using these to map  (\ref{ham1}) into the form
\beq\lb{h_1}
H_1={P_1^2\over2m_1}+{P_2^2\over2m_2}+{1\over2}\left(D_1X_1^2+D_2X_2^2+D_3X_1X_2\right)+
{1\over2}\left(\om_1 P_2X_1-{\om_2}P_1X_2\right)
\eeq
where the new constants $D_1$, $D_2$ and $D_3$ read as
\beq
D_1(B)=C_1+m_2\om_2^2, \qquad D_2(B)=C_2+m_1\om_1^2, \qquad D_3=C_3
\eeq
with the cyclotron frequencies
\beq
\om_{1c}={eB\over m_1c}, \qquad \om_{2c}={eB\over m_2c}.
\eeq
Clearly, by
comparing and forgetting about different involved constants, we notice that the third term
makes difference between (\ref{ham0}) and  (\ref{h_1}). This in fact will play a crucial role in
the forthcoming analysis and allow us to derive different results.

%This Hamiltonian form can be simplified further to get an angular momenta
%operator.

It is convenient to introduce new phase space variables, which can be done
by rescaling those appearing in (\ref{h_1}). Indeed, one can define
the positions as
\beq
x_1=\left({m_1\over m_2}\right)^{{1\over4}}X_1,\qquad x_2=\left({m_2\over m_1}\right)^{{1\over4}}X_2
\eeq
which obviously lead to the momenta
\beq
p_1=\left({m_2\over m_1}\right)^{{1\over4}}P_1,\qquad p_2=\left({m_1\over m_2}\right)^{{1\over4}}P_2.
\eeq
Replacing all, we show that (\ref{h_1}) becomes
\beq\lb{c}
H_2={1\over2m}(p_1^2+p_2^2)+{1\over2}\left(d_1x_1^2+d_2x_2^2+d_3x_1x_2\right)+{\om_c\over2}\left(
x_1p_2-x_2p_1\right)
\eeq
where we have set different constants as
\beq
d_1(B)=D_1\left({m_2\over
  m_1}\right)^{{1\over2}},\qquad d_2(B)=D_2\left({m_1\over
  m_2}\right)^{{1\over2}},\qquad d_3=D_3
\eeq
with unique mass $m=(m_1m_2)^{{1\over2}}$ and the cyclotron frequency
$\om_c=(w_1\om_2)^{1\over2}={eB\over mc}$.
Consequently, (\ref{c}) is showing up an extra term, which is nothing
but the angular momenta and the first is similar to (\ref{ham0}). Therefore,
it will be of interest to deal with such system and underline its
physical properties.

%%%%%%%%%%%%%%%%%%%%%%%%%%%%%%%%%%%%%%%%%%%%
\subsection{Unitary transformation}
%%%%%%%%%%%%%%%%%%%%%%%%%%%%%%%%%%%%%%%%%%%%

According to the expression form (\ref{c}), it appears that
getting the energy spectrum solutions is not a easy task. However, we
can overcome such difficulties by adopting an appropriate approach.
More precisely,
%To get the spectrum as well as the corresponding eigenstates,
we proceed
by making use of an unitary transformation, such that new phase space variables
can be defined by
\beq\lb{d}
y_a=M_{ab}x_b, \qquad \tilde{p_a}=M_{ab}\tilde{p_b}
\eeq
where the matrix $M_{ab}$
\begin{equation}
M_{ab}= \left(%
\begin{array}{cc}
  \cos{\te\over2} & -\sin{\te\over2} \\
   \sin{\te\over2} &  \cos{\te\over2} \\
\end{array}%
\right)
\end{equation}
is an unitary rotation with the mixing angle $\te$. Inserting the
mapping (\ref{d}) into (\ref{c}), one realizes that  $\te$ should
satisfy the condition
\beq\lb{tan}
\tan\te={d_3\over d_2-d_1}
\eeq
to end up with a factorizing Hamiltonian. It is
\beq\label{ham3}
H_3={1\over2m}\left(\tilde{p_1}^2+\tilde{p_2}^2\right)+{k\over2}\left(e^{2\al}
y_1^2+e^{-2\al}y_2^2\right)+{\om_c\over2}\left(y_1\tilde{p_2}-y_2\tilde{p_1}\right)
\eeq
where $k$ and $\al$ are given by
\beq
k=\sq{d_1d_2-{d_3^2\over4}},\qquad e^{\al}={d_1+d_2+\sq{(d_1-d_2)^2+d_3^2}\over2k}
\eeq
%It obvious that
and the condition $4d_1d_2>d_3^2$ must be fulfilled. Note that,
$H_3$ has a form similar to two--dimensional Landau Hamiltonian in the symmetric
gauge. Obviously, they coincide in the case of without coupling, namely $\al=0$.

Before proceeding further, we conclude by
citing some interesting remarks. In doing so, let us return to $H_3$
and define
two operators as
\beq
H_0={1\over2m}\left(\tilde{p_1}^2+\tilde{p_2}^2\right)+{k\over2}\left(e^{2\al}
y_1^2+e^{-2\al}y_2^2\right),\qquad L_3=\left(y_1\tilde{p_2}-y_2\tilde{p_1}\right)
\eeq
where $H_0$ also can be separated into two commuting parts
\beq
{\cal H}_1={1\over2m}e^{-\al}\tilde{p_1}^2+{k\over2}e^{\al}y_1^2, \qquad
{\cal H}_2={1\over2m}e^{\al}\tilde{p_2}^2+{k\over2}e^{-\al}y_2^2.
\eeq
Firstly, one can see that the decoupled Hamiltonian
\begin{equation}\lb{e}
{\cal H}_0 = {1\ov 2m}\ti{p_1}^{2} +
{k\ov 2} y^{2}_{1} + {1\ov 2m}\ti{p_2}^2 +
{k\ov 2}y^{2}_{2}
\end{equation}
can be recovered by taking $\al=0$,  which corresponds to the solution $d_1=d_2$
and $d_3=0$.
Secondly, it is interesting to note that~(\ref{e}) can be
derived by a canonical transformation only from
\begin{equation}
\label{HAM6} {\cal H} = {\cal H}_1 + {\cal H}_2
\end{equation}
as it is pointed out in \cite{kim} and subsequently in \cite{jellal}.
%which is tells us that it is convenient to study $ {\cal H}$ instead of $H_3$.
%Therefore, in the next we focus on this latter and proceed by
%introducing the annihilation and creation operators to solve the
%eigenvalue problem.

According to the above statements, we can rearrange $H_3$  in an
appropriate form. This is
\begin{equation}
H_3 = e^{-\al}{\cal H}_1 + e^{\al}{\cal H}_2 + L_3
\end{equation}
which %is the form that
will be used to tackle different issues in the forthcoming analysis
and in particular the magnetization of the present system.
%As it will
%be clear
This mapping will be helpful in sense that the corresponding
energy spectrum solutions can easily be obtained as we will see soon.

%%%%%%%%%%%%%%%%%%%%%%%%%%%%%%%%%%%%%%%%
\section{Energy spectrum}
%%%%%%%%%%%%%%%%%%%%%%%%%%%%%%%%%%%%%%%%%

As far as
%To explicitly determine
the eigenvalues and  eigenstates are concerned, we adopt
an algebraic method based on different operators in terms of the
phase space ones. This will allow us to obtain the solutions
%After getting %This will us to end up with
%the energy spectrum, we
and investigate their underlying properties.
%by filling the shells with fermions.

%%%%%%%%%%%%%%%%%%%%%%%%%%%%%%%%%%%%%%%%%%%%
\subsection{Algebraic analysis}
%%%%%%%%%%%%%%%%%%%%%%%%%%%%%%%%%%%%%%%%%%%%

It is clear that ${\cal H}$ is a Hamiltonian of two decoupled
harmonic oscillators. Thus it can simply be diagonalized
by defining a set of creation and annihilation operators. They are
given by
\begin{equation}
\lb{CRAN}
a_i = \sqrt{k\ov 2\hb\om} e^{\al\ov 2}y_{i} +
{i \ov \sqrt{2m\hb\om}} e^{-{\al\ov 2}} \tilde{p_i}, \qquad a_i^{\da} =
\sqrt{k\ov 2\hb\om} e^{\al\ov 2}y_{i} - {i \ov \sqrt{2m\hb\om}}
e^{-{\al\ov 2}} \tilde{p_i}
\end{equation}
where the new frequency is
\begin{equation}
\om(B)=\left({4d_1d_2 -d_3^2 \over 4m^2} \right)^{1\over 4}=\sqrt{k\ov m}.
\end{equation}
They satisfy the usual commutation relations
\begin{equation}
 [a_i, a_j^{\dag}] = \delta_{ij}
\end{equation}
and obviously other commutators vanish. It is easy to show that ${\cal H}$ can be mapped
 in terms of
$a_i$ and $a_i^{\dag}$ as
\beq\label{h0map}
{\cal H} = \hb\om \left(a_1^{\dag}a_1 + a_2^{\dag}a_2 +
1\right).
\eeq

According to (\ref{h0map}), it is not hard to
derive the corresponding energy spectrum solutions.
This can be done by solving the eigenvalue equation
\begin{equation}
 {\cal H}|n_1, n_2,\al\rangle = {\cal E}_{n_1,n_2} |n_1,
n_2,\al\rangle
\end{equation}
to get the corresponding states
\begin{equation}
|n_1,n_2,\al\rangle= {(a_1^{\dag})^{n_1} (a_2^{\dag})^{n_2} \ov
\sqrt{n_1!n_2!}} |0, 0,\al\rangle
\end{equation}
as well as the energy spectrum
\begin{equation}
 {\cal E}_{n_1,n_2} = \hb\om \left( n_1 + n_2+ 1\right).
\end{equation}

Due the fact that there are mappings between different Hamiltonian's, one can
build other solutions. In particular,
the spectrum of $H_0$ can easily be deduced from above as
\begin{equation}
 E_{0,n_1,n_2} = {\hb\om} \left[e^{\al} \left(n_1
+{1\ov 2}\right) +e^{-\al} \left(n_2 +{1\ov 2}\right)\right].
\end{equation}
To get that for $H_3$, we need to diagonalize the angular momentum.
In doing so, we define two sets of operators where the first one is
\beq
a_g={1\ov\sq2}\left(a_1+ia_2\right), \qquad a_g^{\da}={1\ov\sq2}\left(a_1^{\da}-ia_2^{\da}\right)
\eeq
and the second reads as
\beq
a_d={1\ov\sq2}\left(a_1-ia_2\right), \qquad a_d^{\da}={1\ov\sq2}\left(a_1^{\da}+ia_2^{\da}\right).
\eeq
They are showing
\beq
\left [a_g,  a_g^{\da}\right] =  \left [a_d,  a_d^{\da}\right]= 1
\eeq
and different commutation relations are nulls. One can note that there
is a conservation of number operators, such as
\begin{eqnarray}
N_1 + N_2= N_d + N_g.
\end{eqnarray}
where we have
%By introducing the number operators
%\begin{eqnarray}
$ N_1 = a_1^{\da}a_1$, $N_2 = a_2^{\da}a_2$,
 $ N_d = a_d^{\da}a_d$ and $N_g = a_g^{\da}a_g$.
%\end{eqnarray}
%At this level,
Now we express the phase space variables
in terms of the new operators to end up with a quantized angular momenta.
This is
%we can write the angular momenta as
\beq
L_3=2\hbar\left(N_d-N_g\right).
\eeq
Obviously, its eigenvalues are $2\hbar\left(n_d-n_g\right)$ and the corresponding eigenvalues
are forming a common basis of $L_3$  and ${\cal H}$.

Finally,  we settled all ingredients to derive
the energy spectrum solutions of $H_3$. Indeed, starting
from the above results, one can see that
(\ref{ham3}) becomes
\beq\label{nham3}
H_{3}=\left(\hbar\om e^\al+\hbar\om_c\right)N_d+\left(\hbar\om e^{-\al}-
\hbar\om_c\right)N_g+\hbar\om\cosh\al.
\eeq
%It can be written in compact form
To write
$H_3$ in compact form, it is convenient to introduce two new frequencies
in terms of the former ones.
These are defined by
\beq
\om_+(B,\al)=\om e^{\al}+\om_c, \qquad \om_-(B,\al)=\om e^{-\al}-\om_c.
\eeq
They are showing a strong dependence to $\al$ and therefore generalize the
standard results~\cite{Gazeau}. Now returning to
map $H_3$  as
%it is easy to obatin the compact form
\beq\lb{H_3}
H_3=\hbar\left(\om_+N_d+\om_-N_g+\om\cosh\al\right).
\eeq

Solving the eigenvalue equation, we can easily derive the energy spectrum solutions. Thus,
the eigenvalues take the form
\beq\label{enre3}
E_{3,n_d,n_g}=\hbar\left(\om_+n_d+\om_-n_g+\om\cosh\al\right), \qquad
n_g, n_d= 0,1,2, \cdots
\eeq
and the eigenstates are given by
\beq
|n_d,n_g,\al\rangle={\left(a_d^{\dag}\right)^{n_d}\left(a_g^{\dag}\right)^{n_g}\over\sq{n_d!n_g!}}|0,0,\al\rangle.
\eeq
%where the groundstate wavefunction is given by
%\begin{equation}
 %\Psi_{0,0,\al} =
%\end{equation}
It is clear that the results obtained so far are $\al$-dependent. This in fact
makes difference with respect to the standard results obtained
by analyzing the Fock--Darwin Hamiltonian~\cite{Gazeau}, which
obviously can
be recovered by setting $\al=0$. % for instance see~\cite{Gazeau}.
At this stage, one may ask about the relevance of such coupling parameter and the
answer will be given in the forthcoming sections where interesting results will be
derived and different discussions will be given.

%%%%%%%%%%%%%%%%%%%%%%%%%%%%%%%%%%%%%%%%%%%%%%%%%%%
\subsection{Underlying properties}
%%%%%%%%%%%%%%%%%%%%%%%%%%%%%%%%%%%%%%%%%%%%%%%%%%%

In investigating the underlying symmetry of the system, one can study the properties of quantum
numbers pairs $(n_d, n_g)$. However, these may not provide simple hints on the ordering
of the energy
$E_{\al,n_d, n_g}$ with the exception of four limiting cases related to the nature of
the coupling parameter and the magnetic field.

%%%%%%%%%%%%%%%%%%%%%%%%%%%%%%%%%%%%
\subsubsection{ Weak coupling case}
%%%%%%%%%%%%%%%%%%%%%%%%%%%%%%%%%%%%%%

To characterize the system behavior, we consider the first case
that corresponds to the limit $\alpha\rightarrow 0$, which means that
the coupling is not strong enough between two oscillators.
This is the case
%the case of
for some physical phenomena. Therefore, we can make
different approximations to approach our findings to well-know
and significant results.

%We start by treating the first case that is the week coupling
%where two limits of magnetic field will be discussed. Indeed,
By taking the limit
%The present situation is equivalent to consider the limit
$\alpha\rightarrow 0$ and
after a simple calculation, we show that the energy spectrum can be
approximated by
\beq\label{aped}
E_{3,n_d,n_g}|_{\alpha\rightarrow 0} \approx \hbar\left[\om(n_d+n_g)+(\al\om+\om_c)(n_d-n_g)+\om\right].
\eeq
We can bring this to an appropriate form by defining
 new quantum numbers. They are
\beq
\lambda=\frac{n_d+n_g}{2}, \qquad \xi=\frac{n_d-n_g}{2}.
\eeq
Thus, one can rearrange (\ref{aped}) as
%\beq E_{n_d n_g} \approx \hbar \om
%\left(2\lambda+2(\al+{\om_c\ov\om})\xi + 1\right). \eeq note that
%this last equation write by
\beq\label{aped2}
E_{\al,\lam,\xi}|_{\alpha\rightarrow 0} \approx 2\hbar
\left[\om\lambda+(\al\om+\om_c)\xi + \frac{\om}{2}\right].
\eeq
It can be identified to the eigenvalues of the Fock--Darwin
Hamiltonian, which can be obtained from (\ref{enre3})
by taking $\al=0$. They are
 %of frequencies $\om$ and $(\al\om+\om_c)$, which corresponds
%to $\al=0$ in (\ref{enre3}). This is
 \beq\label{enre30}
E_{3,n_d,n_g}|_{\al=0}=\hbar\left(\om_+n_d+\om_-n_g+\om\right).
\eeq
Now observing that the following correspondence $(\om_+, \om_-)\lga (\om, \al\om+\om_c)$.
This tells us that (\ref{aped2})
 can be used to analyze the thermodynamical properties in similar way
to that has been done in~\cite{Gazeau}. Moreover, it shows how one can
generalize the Fock--Darwin
Hamiltonian to another one where the interaction still surviving.
%more detail can be found in~\cite{Gazeau}.

At this stage, we can further discuss (\ref{aped2}) by
inspecting two other limits in terms of the field.
%Now let us examine two interesting cases.
In doing so, we suppose that
the cyclotron frequency is much smaller than the
frequency $\om$, i.e. $\om_c\ll \om$, thus we have
\beq
E_{\al,\lam, \xi}|_{\alpha,B\rightarrow 0,} \approx 2\hbar\om
\left(\lambda+\al\xi + \frac{1}{2}\right).
\eeq
According to this, two conclusion can be deduced here. Indeed, firstly we still
have a generalized Fock--Darwin
Hamiltonian but its frequencies are changed now to $(\om,\al\om)$.
Secondly, without coupling we recovers one-dimensional harmonic oscillator of
eigenvalues
%whose eigenvalues are given by
\beq
E_{\lam}|_{\alpha= 0,B\rightarrow 0} \approx 2\hbar\om
\left(\lambda+ \frac{1}{2}\right)
\eeq
whose frequency is $\om=\sqrt{\frac{d_1}{m}}$, which means that we are
in the conditions $d_1=d_2$ and $d_3=0$.

Now, let us treat the second consideration that is the strong
magnetic filed case. In fact, this equivalent to
%However, in the case where
$\om_c\gg \om$ and leads
 \beq E_{\xi}|_{\alpha\rightarrow 0,B\rightarrow\infty,} \approx 2\hbar\om_c \xi.
\eeq
It can be interpreted as
 the squared energy spectrum of the massless Dirac fermions in graphene
under an uniform magnetic field. Fore more detail,
we cite for instance~\cite{jellal2} and reference
therein.

%%%%%%%%%%%%%%%%%%%%%%%%%%%%%%%%%%%%%%%%%%%%%
\subsubsection{Strong coupling case}
%%%%%%%%%%%%%%%%%%%%%%%%%%%%%%%%%%%%%%%%%%%%%

It is immediate and natural to ask about what happens
if the coupling is strong enough and the corresponding
limit cases of magnetic field.

The above inquiry can be answered by
examining the limit $\al\rightarrow\infty$. Thus, returning
to (\ref{enre3}) to show the result
\beq\lb{46}
E_{\al,n_d,n_g}|_{\alpha\rightarrow\infty} \approx \hbar\left[ \om
e^{\alpha}\left(n_d+{1\ov2}\right)+\om_c\left(n_d-n_g\right)\right].
\eeq
Again this can be approximated further by taking other limits.
Indeed, focusing on the case $\om_c\ll \om$ or  $\om_c\ll \om e^{\alpha}$,
it is straightforward to obtain
%(\ref{46}) become
\beq \lb{ff}
E_{n_d} \approx
\hbar\om\left(n_d+{1\ov2}\right)e^{\alpha} \eeq
which is the energy spectrum of harmonic oscillator of
frequency $\om e^{\alpha}$. However for $\om_c\gg \om$,
there is nothing to say and therefore
(\ref{46}) remains as it is because we can not make
 comparison.

In summary, according to the above results we conclude that the
 coupling parameter $\al$ is interesting
parameter of the present theory. In fact,
it can be adjusted to recover different %results and therefore some
models those used to deal with different issues in physics.

%%%%%%%%%%%%%%%%%%%%%%%%%%%%%%%%%%%%%%%%%
\section{Realizing the coherent states}
%%%%%%%%%%%%%%%%%%%%%%%%%%%%%%%%%%%%%%%%%%%

The forthcoming analysis requires a powerful
tools. More precisely, one way to determine the thermodynamical
potential is to use the coherent states approach. Thus, for
the neediness, we follow the standard method to
realize them in terms our language and show their dependence to
the coupling parameter. In fact, we will use
the same steps traced in~\cite{Gazeau}.

%For the neediness, we realize the coherent states
%in terms of the obtained eigenstates for the present system. In fact,
%they can be used to deal with different issues and
%in particular explicitly determine the thermodynamical
%potential.
%To deal with different issues, we need to establish some tools in
%terms of our language. More precisely, we give realization of the
%coherent states for the present system.

The fact that the eigenstates issued from the algebraic method are just
tensor products of Fock harmonic oscillator eigenstates allows one to easily
construct the corresponding coherent states. Indeed, in a standard way, we
have
\begin{eqnarray}
\nonumber
\mid z_d,z_g,\al\rangle\equiv\mid z_d\,\rangle\, \ot\mid z_g\,\rangle =
\exp{\left[ -\frac{1}{2}\left(\vert z_d \vert^2 + \vert z_g \vert^2\right)\right] }\,
\sum_{n_d, n_g}\frac{z_d^{n_d}}{\sqrt{n_d!}}\frac{ z_g^{n_g}}{\sqrt{n_g!}}
\mid n_d, n_g,\al\rangle.
\end{eqnarray}
In terms of the creations operators, we have
%It is can be written as
\begin{eqnarray}
\mid z_d,z_g,\al\rangle= \exp{\left[-\frac{1}{2}\left(\vert z_d \vert^2+\vert z_g\vert^2\right)\right] }\,
 e^{z_d a_d^{\dagger} + z_g a_g^{\dagger}} \mid 0, 0,\al \rangle .
\end{eqnarray}

The above normalized states, should  obey some of the usual properties. Indeed, it
is easy to verify the eigenvector property, such as
\beq
a_d, \mid z_d, z_g,\al \rangle = z_d \mid z_d, z_g,\al \rangle,\qquad
a_g \mid z_d, z_g,\al \rangle = z_g \mid z_d, z_g,\al \rangle.
\eeq
As far as the action identity is concerned, one can obtain the relation
\beq\label{3.2}
\check{ H_3}(z_d, z_g,\al)
\equiv\langle z_d, z_g,\al \mid H_3\mid z_d, z_g,\al \rangle
=\hbar\left(\om_+\vert z_d\vert^2+\om_-\vert z_g\vert^2+\om\cosh\al\right)
\eeq
where
the function $\check{H_3}(z_d, z_g,\al)$ is called lower
 symbol of the operator $H_3$.
It will plays an important role in the present context. The resolution
of the identity reads as
\beq
\mathbb{I}={1\over\pi^2}\int_{\mathbb{C}^2}\mid z_d,z_g,\al\rangle\langle z_d, z_g,\al \mid\,
          d^2z_d \, d^2z_g.
\eeq
where the last property is also crucial in our context.

For any observable $A$ with suitable operator properties (traceclass, $\cdots$),
there exists a unique upper (or covariant) symbol $\hat{A}(z_d, z_g)$ defined by
\beq
A={1\over\pi^2}\int_{\mathbb {C}^2}\, \hat{A}(z_d, z_g,\al)\,
\mid z_d, z_g,\al\rangle\langle z_d, z_g,\al \mid\, d^2z_d \, d^2z_g.
\eeq
As a straightforward illustration, we consider
the upper symbols for the number operators. Hence, one can show
\beq
\hat{N_d}(z_d, z_g,\al) = \vert z_d \vert^2 -1,\qquad
\hat{N_g}(z_d, z_g,\al) = \vert z_g \vert^2 -1.
\eeq
Clearly,  the upper symbol for our Hamiltonian (\ref{H_3}) takes the form
\beq\label{3.6}
\hat{ H_3}(z_d, z_g,\al)=\hbar\left(\om_+\vert z_d \vert^2+\om_-\vert z_g
\vert^2 -\om\cosh\al\right).
\eeq

To setup all what we need for our task, we recall
an useful trace identity for a given traceclass
observable $A$. This is
\beq
\Tr A ={1\over\pi^2}\int_{\mathbb C^2}\, \check{A}(z_d,z_g,\al)\, d^2z_d\, d^2z_g
={1\over\pi^2}\int_{\mathbb C^2}\, \hat{A}(z_d, z_g,\al)\, d^2z_d\, d^2z_g
\eeq
where the symbol function $\check{A}$ is
\beq
\check{A}(z_d, z_g,\al)\equiv\langle z_d, z_g,\al\mid  A\mid z_d, z_g,\al\rangle.
\eeq

We close this part by noting that all involved quantities are $\al$-dependent. Obviously,
the standard results can be recovered by switching off the coupling
parameter~\cite{Gazeau}. On the other hand, we will see how the above materials can be employed to
deal with different issues and in particular determine
the thermodynamical potential. This will be done by adopting two methods, which concern
the Berezin--Lieb inequalities
and Fermi--Dirac trace formulas.

%%%%%%%%%%%%%%%%%%%%%%%%%%%%%%%%%%%%%%%%%%%%%%%%%%%%%%%%%%%%%%%%%%
\section{Berezin--Lieb inequalities} %for thermodynamical potential}
%%%%%%%%%%%%%%%%%%%%%%%%%%%%%%%%%%%%%%%%%%%%%%%%%%%%%%%%%%%%%%%%%%

Having derived and settled all necessary tools, we now show that how
they can be used to study the magnetism of the system under consideration.
%In the beginning,
In doing so,
we start by defining the physical quantities those
will be discussed in the present context. One way to do so
is to evaluate  the thermodynamical potential, %through
%which can be achieved by an
which can be done, in the first stage,
by adopting %an
%appropriate method
the Berezin--Lieb inequalities. Subsequently, we
treat the asymptotic behavior of the obtained results by considering the liming cases
of the coupling parameter.

%%%%%%%%%%%%%%%%%%%%%%%%%%%%%%%%%%%%%%%%%%%%%%%%%%%%%%%%%%%%%%%%%%
\subsection{Physical quantities} %for thermodynamical potential}
%%%%%%%%%%%%%%%%%%%%%%%%%%%%%%%%%%%%%%%%%%%%%%%%%%%%%%%%%%%%%%%%%%

%Let us now enter the core of the physical question we addressed in the
%introduction.

The magnetism of the model under hand can be investigated
by adopting the standard method of statistical mechanics. This will be done
by making use of different approximations to simplify our problem. In fact,
we begin by assuming that the total number $\langle N_e \rangle$ of electrons is large
enough for making no appreciable difference between a grand canonical ensemble
and a canonical one.

On the light of the above considerations and obtained results, we proceed by using
the magnetic moment $M$ definition. This is
%is defined by
\beq\label{4.1}
M = -\left(\frac{\pa \Om}{\pa H}\right)_{\mu}
\eeq
where thermodynamical potential $\Om$ can be obtained from the partition function. In terms of
our model, it is
\beq\label{4.2}
\Om = -\frac{1}{\be}\Tr\log{\left[1+\exp\{-\be(H_3-\mu)\}\right]}
\eeq
as usual we have set  $\be = 1/(k_BT)$. According to (\ref{4.1}) and (\ref{4.2}), we show the result
\begin{equation}\label{4.4}
M = -2\mu_B\, \Tr\left(\frac{N_d-N_g} {1+ \exp\{\be(H_3-\mu)\}}\right).
\end{equation}
%which gives the quantity
Replacing $H_3$ by its expression and tracing to end up with
%After tracing and replacing, we find
\begin{equation}
M = -2\mu_B\, \sum_{n_d,n_g=0}^{\infty} %\sum_{n_g=0}^{\infty}
\frac{n_d-n_g}
     {1+\kappa_{\pm}^{-1} \exp\{\be\hbar(\om_+n_d+\om_-n_g)\}}
\end{equation}
where $\mu_B =\hbar e/(2mc)$ is the Bohr magneton and $\ka_{\pm}$
are given by
\beq\lb{4.5}
\ka_{\pm}(B,T,\al)=\exp{\left[\be(\mu \pm \hbar
  \om\cosh\al)\right]}.
\eeq

On the other hand, the average number of electrons
can be evaluated by introducing the Fermi distribution function.
That is
\beq\label{disf}
f(E) = \frac{1} {1 + \exp\{\be (E - \mu)\}}.
\eeq
Therefore, in our case we have
\beq
\langle N_e\rangle=\sum_{n_d,n_g=0}^{\infty} %\sum_{n_g=0}^{\infty}
f(E_{3,n_d,n_g})
= \mbox{Tr}f({H_3}) = -\partial_{\mu} \Om
\eeq
which is showing that there are two possibilities to get $\langle N_e\rangle$
either summing all distributions (\ref{disf}) or deriving $\Om$ with
respect to chemical potential.
%At this level, it is clear that
Clearly,
to go further in
evaluating different physical quantities, one should explicitly
determine $\Om$. %This can be done by adopting different approaches.
%One way to do
%so is to use an approach of
%This can be achieved by making use a method based on
%the Berezin--Lieb inequalities.
%To evaluate different thermodynamical quantities, we need to determine
%$\Om$. In doing so, we start by using the Berezin-Lieb inequalities.

%%%%%%%%%%%%%%%%%%%%%%%%%%%%%%%%%%%%%%%%%%%%%%%%%%%%%%%%%%%%%%%%%%%%%%%%%%%%
\subsection{Calculating the thermodynamical potential}
%%%%%%%%%%%%%%%%%%%%%%%%%%%%%%%%%%%%%%%%%%%%%%%%%%%%%%%%%%%%%%%%%%%%%%%%%%%%%%

The thermodynamical potential is very much needed
to describe the quasi-classical behavior of present system.
This can be calculated by adopting some technical methods
like for instance
the Berezin--Lieb inequalities. In fact,
it is based on some general statement that is
%In this subsection we can apply the Berezin-Lieb inequalities to
%describe the quasi-classical behavior of the thermodynamical
%potential.  These inequalities say that,
for any convex function $g(A)$ of the observable
$A$, one can write the inequalities
\beq\label{ineq}
{1\over\pi^2}\int_{\mathbb C^2}g(\check{A})\,d^2z_d\,d^2z_g
\leq \Tr g(A) \leq
{1\over\pi^2}\int_{\mathbb C^2}g(\hat{A}) \, d^2z_d\,d^2z_g
\eeq
where the lower and upper symbol functions  $(\check{A}, \hat{A})$ are defined
before. This tells us that knowing inferior and superior boundaries of
a given observable, one can derive its trace.

At this level, we have all ingredients needed to do our task. Indeed,
%to determine $\Om$. Therefore,
 an straightforward application
of (\ref{ineq}) gives the result
%Now we have all materials to do our task. Indeed, an straightforward application
%of (\ref{ineq}) to the thermodynamical potential gives the result
\begin{eqnarray}\lb{4.6}
-{1\over\be\pi^2}\int_{\mathbb C^2}\log{\left[1+\exp\{-\be(\hat{ H_3}-\mu)\}\right]}\,
d^2z_d \, d^2z_g \leq \Om \nonumber\\
\leq
-{1\over\be\pi^2}\int_{\mathbb C^2}\log{\left[1+\exp\{-\be(\check{H_3}-\mu)\}\right]}\,
d^2z_d \, d^2z_g.
\end{eqnarray}
%Inserting,
After mapping
(\ref{3.2}) and (\ref{3.6}) into (\ref{4.6}),  we end up with
\begin{eqnarray}\label{intp}
\nonumber
-{1\over\be}\int_{0}^{\infty}du_d\,\int_{0}^{\infty} du_g\,
\log\left[1+\exp\{-\be(\hbar\om_+u_d+\hbar\om_-u_g-\hbar\om\cosh\al-\mu)\}\right]\leq\Om\\
 \leq -{1\over\be}\int_{0}^{\infty}du_d\,\int_{0}^{\infty}du_g\,
\log\left[1+\exp\left\{-\be(\hbar\om_+u_d +\hbar\om_- u_g+\hbar\om\cosh\al-\mu)\right\}\right]
\end{eqnarray}
where we have set $u_d = \vert z_d \vert^2$ and $u_g = \vert z_g \vert^2$. The solution
can be obtained by
making some rearrangement followed by an integration. Indeed, by changing
%changing the integration
variables as
%to other ones, such as
\beq
u= \be\hbar(\om_+ u_d +\om_- u_g), \qquad v=\be\hbar\om_+ u_d
\eeq
%as well as performing an integration by part of (\ref{intp}). Indeed,  in terms
we show that,
in terms of the parameters $\ka_{\pm}$ (\ref{4.5}), (\ref{intp}) becomes
\beq \label{4.8}
\Phi(\ka_+)\leq\Om\leq\Phi(\ka_-)
\eeq
where the  function $\Phi$ is given by
\beq\lb{5}
\Phi(\ka)= -{\ka\over2\be(\be\hbar)^2\om_+\om_-}
\int_0^{\infty}\frac{u^2 e^{-u}}{1+\ka e^{-u}}\, du.
\eeq
Actually, the problem of determining $\Om$ is restricted to find the solutions of such integral.
This can be done by defining a new parameter
%This can be intergated by
%setting  $\lambda$ as
\beq
\lambda={\om_+\om_-}={\om^2-\om_c^2-2\om\om_c\sinh\al}
\eeq
%and according to the sign of $\ka$, we obtain
and distinguishing between the sign of $\ka_{\pm}$
to end up with the  solutions %of (\ref{5}) can be written as
\begin{eqnarray}
 \Phi(\ka)= \left\{ \begin{array}{ll}
 {1\over\be\lam(\be\hbar)^2}\,F_3(-\ka), & \qquad \ka \leq 1\\
 {1\over\be\lam(\be \hbar)^2}
\left\lbrack -{(\log{\ka})^3\over6}-{\pi^2\log{\ka}\over6}+ F_3(-\ka^{-1})
\right\rbrack, & \qquad \ka >1
 \end{array}
\right. \label{4.9}
\end{eqnarray}
where we have introduced here the function $F_s$ of the Riemann-Fermi-Dirac
type. For a given variable $z$, it reads as
\beq
F_s(z)=\sum_{m=1}^{\infty} \frac{z^m}{m^s }.
\eeq
Note in passing that $\Phi$ is depending to the sign of $\lam$ as well.
Discussions about such matters will be reported next. On the other hand,
the above results can be discussed by separately considering
the high and low temperature regimes.

%%%%%%%%%%%%%%%%%%%%%%%%%%%%%%%%%%%%%%%%%%%%%%%%%%%%%%%%%%%%%%%%%%%%%%%%
\subsubsection{High temperature regime}
%%%%%%%%%%%%%%%%%%%%%%%%%%%%%%%%%%%%%%%%%%%%%%%%%%%%%%%%%%%%%%%%%%%%%%%%

Having the expression (\ref{4.8}) together with (\ref{4.9}),
one can introduce an appropriate approximation to further simplify the
form of $\Om$ and derive interesting results. 
To achieve this goal, we can analyze two liming cases of
the temperature of the present system. 

%This can be done, in the first stage,
%by focusing
%In fact, let us focus
 We start our analysis by considering the high temperature regime that
 corresponds to the condition $\vert \mu \pm \hbar \om\cosh\al \vert \ll k_B T$.
By taking into account, we find
%$\ka_{\pm} \approx 1$. Then the thermodynamical potential is
%approximately equal to
%Consequentely, one can find
\beq\label{apptp}
\Om \approx \frac{k_BT} {\lam}\left(\frac{k_B T}{\hbar}\right)^2 F_3(-1)
\approx -0.901543\,\frac{k_BT} {\lam}\left(\frac{k_B T}{\hbar}\right)^2.
\eeq
This is a nice form that can be further discussed.
Recall that the involved parameter $\lam$ is magnetic field and $\al$-dependent, which gives
a generalization to the already %then without coupling
%(\ref{apptp}) coincides exactly with that
obtained in~\cite{Gazeau}. Obviously without coupling, they coincide.

On the other hand, one can report different discussions related
to the above form of $\Om$ in terms of the coupling parameter$\al$. With these we can
show what makes difference with respect to the standard case, i.e. $c_1=c_2$ and $c_3=0$. 
By doing this, we can summarize the following results:
%we summarize some results in the following points.
\begin{itemize}
\item By inspecting the form of $\Om $, one can immediately notice the first general result. Indeed,
by considering a negative $\lam$ we end up with a positive $\Om$, which
can not be obtained from the standard results~\cite{Gazeau}.
\item
(\ref{apptp}) is magnetic field dependent as well and therefore
the present case exhibits an magnetism behavior. This statement can
be confirmed by explicitly determining the magnetic moment and susceptibility.
\item
The easiest way to obtain the magnetic moment is that one can require for instance the following
configuration: 
\begin{equation}
 \om\lga \om_c, \qquad \sinh \al =\mbox{\sf finite number}.
\end{equation}
%
%$\om\lga \om_c$ and $\sinh \al$ is a finite number.
\end{itemize}

After giving quick conclusions by looking at the form (\ref{apptp}), now let us
be much more accurate and derive explicit results. Indeed, 
after a straightforward calculation, we show that the magnetic moment takes
the form
\beqar
 M &=& 1.803086\ k_BT \left(\frac{k_BT e}{\lam\hbar mc}\right)^2
\Bigg[ \frac{B}{2\om^2} \left ( \frac{c_2+m_1\om_1^2}{m_1} + \frac{c_1+m_2\om_2^2}{m_2}\right)\nonumber \\
&&
-\sinh\al \left \{\frac{\om mc}{e} + \frac{2B\om_c}{\om^3} 
\left ( \frac{c_2+m_1\om_1^2}{m_1} + \frac{c_1+m_2\om_2^2}{m_2}\right)\right\}
- 1 \Bigg].
\eeqar
This allows us to end up with
the susceptibility
\begin{equation}
\chi= - 3.62172\ k_BT \left(\frac{k_BT e}{\hbar mc}\right)^2 
\left[\frac{8m}{4c_1c_2 -c_3^2} \sinh\al
+\frac{m}{4c_1c_2 -c_3^2}  
\left ( \frac{c_2}{m_1} + \frac{c_1}{m_2}\right)
+1
\right].
\end{equation}
It is clear that $\chi$ is behaving as a linear function in
terms of the hyperbolic function $\sinh\al$. This results
in fact is showing the difference with respect to
the case without coupling where there is no susceptibility and
therefore no effect is obtained at high temperature regime.

%%%%%%%%%%%%%%%%%%%%%%%%%%%%%%%%%%%%%%%%%%%%%%%%%%%%%%%%%%%%%%%%%%%%%%%%
\subsubsection{Low temperature regime}
%%%%%%%%%%%%%%%%%%%%%%%%%%%%%%%%%%%%%%%%%%%%%%%%%%%%%%%%%%%%%%%%%%%%%%%%

To accomplish our analysis in terms of temperature, we discuss
the last case. This can be achieved by considering 
the more realistic case, which is
$\mu \gg \hbar \om\sinh \al$ and $\mu \gg k_B T$. With these, we will be 
able to derive interesting results and deduce different conclusions.

%We acomplish our analysis by discussing the last 
%If we return to the $\Om$ expression, one can notice that there are
%other inspectation can be investigated. Especially, one can address
%to the more realistic case that is
%$\mu \gg \hbar \om/2$ and $\mu \gg k_B T$. 

After considering the above two limiting cases, we show that (\ref{4.9})
can be written as combination of three parts. This is
\beq
\Phi(\ka_{\pm}) = A \mp \frac{\Delta}{2} + S_{\pm} \label{4.11}
\eeq
where different terms are given by
%Taking into account the expression of $\ka_{\pm}$ we can obtain from
%(\ref{4.9}) the three parts of $\Phi$. They are
\begin{eqnarray}\lb{k}
\nonumber
A(B,T,\al) &=&
-{\mu\over 2\lam}\left[{1\over3}\left({\mu\over\hbar}\right)^2 +
{\om^2\cosh^2\al} + {\pi^2\over3}
\left({k_B T\over\hbar}\right)^2\right]\\
{\Delta\over2}(B,T,\al) &=&
{\hbar\om\cosh\al\over 2\lam}\left[\left({\mu\over\hbar}\right)^2+
{1\over3} {\om^2\cosh^2\al} + {\pi^2\over6}
\left({k_BT\over\hbar}\right)^2 \right]\\ \nonumber
S_{\pm} (B,T,\al)&=& \frac{k_B T} {\lam} \left({k_B T\over\hbar}\right)^2
F_3(-\exp{[-\be( \mu \pm \hbar \om\cosh\al )]}).
\end{eqnarray}
According to these functions, we notice that
%These different relations show that
$\Om$ is in the interval
$\left[A+S_+-{\Delta\over2},A+S_-+{\Delta\over 2}\right]$.
This will be used to derive different results in the present context.

We can go further by making an important assumption. In fact, we 
restrict ourselves to the condition  
\begin{equation}
 e^{\pm\hbar\om\cosh\al}\approx 1.
\end{equation}
%$e^{\pm\hbar\om\cosh\al}\approx1$. 
In this situation, one can see
that $S_{\pm}$ is reduced to
\beq
S_{\pm}\approx S_0(B,T,\al)=\frac{k_BT}{\lam}\left({k_BT\over \hbar \om_0}\right)^2F_3
\left(-e^{-\beta\mu}\right). %\left(1+{2\om\om_c\over\om_0^2}\al\right).
\eeq
Moreover, taking into account the above limiting cases, %$\mu \gg \hbar \om/2$ and $\mu \gg k_B T$,
we show
%and noting that the reation
\begin{equation}
\frac{\Delta}{|A+S_0|}\lga 0.
\end{equation}
Combining all to end up
with the form
%we can write $\Om(B,T,\al)$ as
\begin{equation}\label{omapp}
%\Om(B,T,\al)
\Om\,
%&\approx&
%A+S_0\\
 \approx\, \frac{1}{\lam}\,
\left[-{\mu\over2}\left\{ {1\over3}\left({\mu\over\hbar \om_0}\right)^2 +
\om^2\cosh^2\al+{\pi^2\over3}
\left({k_B T\over\hbar \om_0}\right)^2\right\} +
 k_BT\left({k_BT\over \hbar \om_0}\right)^2F_3\left(-e^{-\beta\mu}\right)\right].%\nonumber\\
%&&
%\times
%\left(1+{2\om\om_c\over\om_0^2}\al\right). %.\nonumber
\end{equation}
This in fact can be used to deduce different physical quantities. In particular, we
evaluate the average number of electrons to obtain
\begin{equation}
%\nonumber
%{\cal N}_e(B,T,\al)
\langle N_e\rangle (B,T,\al)
\approx \frac{1}{\lam}
\left({\mu\over\hbar}\right)^2\left[{1\over2}+\left({\hbar\om\over
\mu}\right)^2\right]+{\pi^2\over3}\left({k_BT\over\mu}\right)^2+
\left({k_BT\over\mu}\right)^2F_2\left(-e^{-\beta\mu}\right).
\end{equation}
As far as the magnetic moment is concerned, one can obtained
a complicated form. This is due to the fact $\lam$ and $\om$ are
magnetic field dependents. However, we can get more information
by inspecting some limiting cases. These will also
offer for us a way
to emphasis what makes difference with respect to other approaches
and in particular~\cite{Gazeau}.
%let us study the underlying properties of the above functions.
This can be done
by discussing the nature of
the coupling parameter involved in the game.

%considering the limiting cases of the coupling parameter.
%by considering three limiting cases: week, without and strong couplings.

%%%%%%%%%%%%%%%%%%%%%%%%%%%%%%%%%%%%%%%%%%%%%%
\subsection{Asymptotic behavior}
%%%%%%%%%%%%%%%%%%%%%%%%%%%%%%%%%%%%%%%%%%%%%%%

Having derived a general expression of the thermodynamical potential,
one can ask about further simplifications to characterize
the system behavior in some special cases. More precisely, how the above
results can be approximated by inspecting
the limits:
%
%At this level, one may ask about the relevant of the
%coupling parameter. To reply this inquiry, one may
%distinguish different cases. This concerns three
%different cases: $
 $\al=0$, $\al\ll1$ and $\al\rightarrow\infty$. The reply
of such question
is the subject of the next investigations.

%%%%%%%%%%%%%%%%%%%%%%%%%%%%%%%%%%%%
\subsubsection{ Without coupling } %Without coupling}
%%%%%%%%%%%%%%%%%%%%%%%%%%%%%%%%%%%%%%%%%

It is natural to ask about the case $\al = 0$. To reply this inquiry, one can
return to the former analysis to show that
the different quantities given in (\ref{k}) can be restricted to
the functions
\begin{eqnarray}\label{al=0}
\nonumber
A(B,T,0) &=&
-{\mu\over2}\left[{1\over3}\left({\mu\over\hbar \om_0}\right)^2 +
\left({\om\over\om_0}\right)^2+{\pi^2\over3}
\left({k_B T\over\hbar \om_0}\right)^2\right]\\
{\Delta\over2}(B,T,0) &=&
{\hbar\om\over2}\left[\left({\mu\over\hbar\om_0}\right)^2+
{1\over3}\left({\om\over\om_0}\right)^2+{\pi^2\over6}
\left({k_BT\over\hbar \om_0}\right)^2 \right]\\ \nonumber
S_{\pm}(B,T,0) &=& k_B T \left({k_B T\over\hbar \om_0}\right)^2
F_3(-\exp{[-\be( \mu \pm \hbar \om)]}).
\end{eqnarray}
They show that the thermodynamical potential lies in $\left[A+S_+-{\Delta\over
    2},A+S_++{\Delta\over 2}\right]$. Note that,
these exactly coincide with those obtained by analyzing a confined two-dimensional
system in the presence of an uniform magnetic field~\cite{Gazeau}.

To reproduce most of results derived in~\cite{Gazeau}, one can inspect (\ref{al=0})
by making an approximation. That is $S_{\pm}$ can be replaced by
\beq
S_0=k_BT\left({k_BT\over \hbar \om_0}\right)^2F_3\left(-e^{-\beta\mu}\right)
\eeq
to end up with the form
\begin{equation}
%\nonumber
\Om
%&\approx& A+S_0\\
 \approx
\left[-{\mu\over2}\left\{{1\over3}\left({\mu\over\hbar \om_0}\right)^2 +
\left({\om\over\om_0}\right)^2+{\pi^2\over3}
\left({k_B T\over\hbar \om_0}\right)^2\right\}+
 k_BT\left({k_BT\over \hbar \om_0}\right)^2F_3\left(-e^{-\beta\mu}\right)\right].
\end{equation}
Therefore,
the average number of electrons is given by
\begin{eqnarray}\label{al=0n}
\langle N_e\rangle\approx
{1\ov2}\left[-\left\{\left({\mu\over\hbar \om_0}\right)^2 +
\left({\om\over\om_0}\right)^2+{\pi^2\over3}
\left({k_B T\over\hbar \om_0}\right)^2\right\}+\left({k_BT\over \hbar \om_0}\right)^2F_2\left(-e^{-\be\mu}\right)\right].
\end{eqnarray}
as well as the magnetic moment
\beq\lb{chi}
M\approx4\mu\left({\mu_B\over\hbar\om_0}\right)^2B.
\eeq
The corresponding susceptibility $\chi_p$ read as
\beq
\chi_p= 4\mu\left({\mu_B\over\hbar\om_0}\right)^2.
\eeq

One can also inspect other approximations. Indeed, by requiring that
 $\mu\gg k_BT$ and   $\mu\ll\hbar\om$, we show that (\ref{al=0n})
can be written
\begin{equation}\lb{Ne}
\langle N_e\rangle \approx
{1\over2}\left({\mu\over\hbar\om_0}\right)^2.
\end{equation}
The above derivation show that our results are general in sense that
after making appropriate choices one
can recover already significant published works.
%This summarizes the recovered results already obtained in~\cite{Gazeau}.

%%%%%%%%%%%%%%%%%%%%%%%%%%%%%%%%%%%%%%%%%
\subsubsection{ Week coupling } %$\al\ll1$}
%%%%%%%%%%%%%%%%%%%%%%%%%%%%%%%%%%%%%%%%

We start our analysis by dealing with the first case that corresponds to $\al\ll1$. Clearly,
an expansion of different quantities entering in the game is very much needed. Indeed,
by taking the first order of $\al$, we can approximate
(\ref{omapp}) as
%\begin{eqnarray}
%\nonumber
%A(B,T,\al) &=&
%-{\mu\over2}\left[{1\over3}\left({\mu\over\hbar \om_0}\right)^2 +
%\left({\om\over\om_0}\right)^2+{\pi^2\over3}
%\left({k_B T\over\hbar \om_0}\right)^2\right]\left(1+{2\om\om_c\over\om_0^2}\al\right)\\
%{\Delta\over2}(B,T,\al) &=&
%{\hbar\om\over2}\left[\left({\mu\over\hbar\om_0}\right)^2+
%{1\over3}\left({\om\over\om_0}\right)^2+{\pi^2\over6}
%\left({k_BT\over\hbar \om_0}\right)^2 \right]\left(1+{2\om\om_c\over\om_0^2}\al\right) \\ \nonumber
%S_{\pm}(B,T,\al) &=& k_B T \left({k_B T\over\hbar \om_0}\right)^2
%F_3(-\exp{[-\be( \mu \pm \hbar \om)]})\left(1+{2\om\om_c\over\om_0^2}\al\right)
%\end{eqnarray}
\begin{eqnarray}\label{omappa1}
%\Om(B,T,\al)
\Om\,
%&\approx&
%A+S_0\\
 &\approx&
\left[-{\mu\over2}\left\{ {1\over3}\left({\mu\over\hbar \om_0}\right)^2 +
\left(\frac{\om}{\om_0}\right)^2+{\pi^2\over3}
\left({k_B T\over\hbar \om_0}\right)^2\right\} +
 k_BT\left({k_BT\over \hbar \om_0}\right)^2F_3\left(-e^{-\beta\mu}\right)\right]\nonumber\\
&&
\times
\left(1+{2\om\om_c\over\om_0^2}\al\right). %.\nonumber
\end{eqnarray}
where we have set $\om_0^2= \om^2-\om_c^2$.

Obtaining (\ref{omapp}), it is worthwhile to ask about the related physical
quantities to characterize their behaviors in terms
the coupling parameter for the present case. Using the former definitions
to show that
%Indeed, returning to the definitions cited before, we show that
the average number of electrons is
%\begin{eqnarray}
%{\cal N}&=&
%-\pa_{\mu}\left(A+S_0\right).
%\end{eqnarray}
%This leads to end up with the quantity
\begin{equation}
%\nonumber
%{\cal N}_e(B,T,\al)
\langle N_e\rangle (B,T,\al)
\ \approx\
\left({\mu\over\hbar}\right)^2\left[{1\over2}+\left({\hbar\om\over
\mu}\right)^2\right]+{\pi^2\over3}\left({k_BT\over\mu}\right)^2+
\left({k_BT\over\mu}\right)^2F_2\left(-e^{-\beta\mu}\right) \left(1+{2\om\om_c\over\om_0^2}\al\right).
\end{equation}
On the light of the assumptions
$\mu\gg k_BT$ and  $\mu\gg\hbar\om$,
we obtain
\begin{equation}\lb{33}
\langle N_e\rangle (B,\al) \approx
{1\over2}\left({\mu\over\hbar\om_0}\right)^2
+\om\om_c \left({\mu\over\hbar\om_0^2}\right)^2\al.
\end{equation}
Clearly, the second term in right hand is appearing
a correction to the average number of electrons.
This is agreed by canceling the coupling
to recover the standard result~(\ref{Ne}).

Now let us investigate the magnetism in such case. Indeed,
a straightforward calculation gives
%One can also go straightforwardly to get
 the magnetic moment  as
\beq\lb{moal}
{ M}(B,T,\al) ={2\mu_B\ov\hbar\om_0}\left[\mu\om_c\left(1+{2\om\om_c\ov\om_0^2}\al\right)-
2\left({\om^2+\om_c^2\ov\om}\right)\Om(B,T,0)\right]
\eeq
where $\Om(B,T,0)$ is the thermodynamical potential corresponding
to the standard case, i.e. $\al=0$.
Considering $\mu\gg k_BT$ and  $\mu\gg\hbar\om$, we show
%The above expressions can be simplified further to get linear dependence
%in terms of the coupling parameter.
%More precisely, we introduce other conditions to how clearly
%the asymptotic behavior of the present system with respect to $\al$. Then
%by considering the cases $\mu\gg k_BT$ and  $\mu\gg\hbar\om$,
%we reduce $\Om$ to the new form
%\beq
%\Om\approx -{\mu\over6}\left({\mu\over\hbar \om_0}\right)^2\left(1+
%{2\om\om_c\over\om_0^2}\al\right).
%\eeq
%Consequently, we obtain %$\langle N_e\rangle$ as
%as well as
\beq\lb{moal2}
M(B,\al) \approx {2\ov3}{\mu\mu_B\ov\hbar\om_0^2}\left({\mu\ov\hbar\om_0}
\right)^2\left({\om^2+\om_c^2\ov\om}\right)\al.
\eeq
%To close this part,
One important thing should be noted here is that
the magnetic moment is behaving like a linear function in terms of $\al$.
Obviously, without coupling we end up a null magnetization. This means that, we have
like phase transition from coupling to decoupling system. This
point might be investigated further to deal with other issues in statistical
physics. Furthermore, by carefully identifying (\ref{chi})
to (\ref{moal2}), one can fix $\al$ to reproduce the
orbital paramagnetism. Indeed, the solution can be written as
\begin{equation}
\al = \frac{2\hbar \om_0^2\mu_B\om}{\om^2+ \om^2_c} B.
\end{equation}
On the other hand, one can also make another choice of the coupling parameter
to get interesting result (\ref{chi}). In fact, here also one can reproduce
the Landau diamagnetism.

%%%%%%%%%%%%%%%%%%%%%%%%%%%%%%%%%%%%%%%%%%%%%%
\subsubsection {Strong coupling}
%%%%%%%%%%%%%%%%%%%%%%%%%%%%%%%%%%%%%%%%%%%%%%

To complete our analysis we consider the last case
that is the strong coupling limit.
This of course will shine light on the system behavior
at such case and therefore allow us to get more interesting
results. To clarify this, we take
%Then talking
%To end the present investigation, we deal with the last case. In fact,
%in the limiting case
the limit
$\al\lga \infty$  to obtain
%\begin{eqnarray}
%\nonumber
%A &=&
%-{\mu\over8}\left({1\over \om_c}\right)^2\\
%{\Delta\over2} &=&
%{\hbar\om\over12}\left( {1\over \om_c}\right)^2e^{\al}\\ \nonumber
%S_{\pm} &=& 0 .
 %\end{eqnarray}
\begin{equation}
A =
-{\mu\over8}\left({1\over \om_c}\right)^2, \qquad
{\Delta\over2} =
{\hbar\om\over12}\left( {1\over \om_c}\right)^2e^{\al}, \qquad
S_{\pm} = 0
 \end{equation}
which leads to the thermodynamical potential
\begin{equation}
\Om
 \approx -{\mu\over8}\left({1\over \om_c}\right)^2.
\end{equation}
It is clear that
 the average number reads as
\begin{equation}\label{nbf}
\langle N_e\rangle \approx
%{1\over8}\left({1\over\hbar\om_c}\right)^2
{1\over8}\left({mc\over\hbar
eB}\right)^2
\end{equation}
which behaves as  the inverse of magnetic field. It seems that (\ref{nbf})
is sharing some common features with
%results can be interpreted in terms of
the quantum Hall effect results~\cite{prange}.
Indeed, one has to recall that the filling factor is defined as
the ration between $\langle N_e\rangle $ and the quantized flux.
More precisely, we can write
\beq
\frac{\langle N_e\rangle }{N_{\phi}}, \qquad N_{\phi}=\frac{BS}{\phi_0}
\eeq
where $S$ is the system area and $\phi_0=\frac{he}{c}$. Clearly, we can adjust
all parameter to show that effectively we have something related to
the quantum Hall effect.

In summary, the Berezin--Lieb inequalities are a powerful tools
one can use to study the thermodynamical behavior
for a given system. As we have seen so far, a straightforward
application of such approach allows us to
derive different interesting results.
On the other hand, as we claimed before there is another way to
do so and this will be tackled next.

%%%%%%%%%%%%%%%%%%%%%%%%%%%%%%%%%%%%%%%%%%%%%%%%%%%%%%%%%%%%%%%%%%%%%%%%%%%%%%%%%%
\section{Fermi--Dirac trace formulas}%Thermodynamical properties}
%%%%%%%%%%%%%%%%%%%%%%%%%%%%%%%%%%%%%%%%%%%%%%%%%%%%%%%%%%%%%%%%%%%%%%%%%%%%%%%%%%%%

As we claimed before, we use the second method to explicitly determine
the exact expressions for  the thermodynamical potential. This is in fact
based on the Fermi--Dirac trace formulas, which does not include
include approximations in the derivation of $\Om$ and therefore
makes difference with respect to the Berezin--Lieb inequalities.
Subsequently, we restrict ourselves to the evaluation of the
average number of electrons and magnetic moment as well as their
expressions at zero coupling.

%In this section will see how the second method can be applied to
%determine the thermodynamical potential. This is in fact based on
%the Fermi--Dirac trace formulas.

%%%%%%%%%%%%%%%%%%%%%%%%%%%%%%%%%%%%%%%%%%%%%%%%%%%%%%%%%%%%%%%%%%%%%%%%%%%%%%%%%%
\subsection{Exact expressions of $\Om$}
%%%%%%%%%%%%%%%%%%%%%%%%%%%%%%%%%%%%%%%%%%%%%%%%%%%%%%%%%%%%%%%%%%%%%%%%%%%%%%%%%%%%

Using the machinery developed in the reference~\cite{Gazeau}, we can
derive an exact form of $\Om$. This can be done by making
an straightforward application of the Fermi--Dirac trace formulas,
in particular (\ref{a.5}) and (\ref{a.7}) in the appendix. Using
 (\ref{a.6}) to define a function $\Theta(k)$ in terms of language
as
\beq
\Theta(k)=\Tr\left[e^{-(ik+1){\beta\over2}H_3}\right]
\eeq
where the Hamiltonian $H_3$ is given in~(\ref{ham3}).
%\beq
%H_3=\hbar\left(\om_+N_d+ \om_-N_g+ \om\cosh\al\right).
%\eeq
After
replacing $H_3$, we end up with
\beq
\Theta(k)=\Tr\left[e^{-(ik+1){\hbar\beta\over2}\left(\om_+N_d+ \om_-N_g+ \om\cosh\al\right)}\right].
\eeq
which can be written as
\beq
\Theta(k) =  e^{-(ik+ 1)\frac{\be}{2}\hbar\om\cosh\al}
 \,
\frac{1}{1-e^{-(ik+ 1)\frac{\be}{2}\hbar \om_+}}
\,
\frac{1}{1 - e^{-(ik + 1)\frac{\be}{2}\hbar \om_-}}.
\eeq

Now let tackle our problem by writing
the Fourier integral representation for the thermodynamical potential. This is
\beq
\Om = -\frac{1}{\be }\int_{-\infty}^{+\infty}
\frac{e^{-(ik+ 1)\frac{\be}{2}(\hbar\om\cosh\al-\mu)}}
     {2\cosh{\frac{\pi}{2}k}} \,
\left(\frac{1}{ik+1}\right)
\left(\frac{1}{1-e^{-(ik+ 1)\frac{\be}{2}\hbar \om_+}}\right)
\left(\frac{1}{1 - e^{-(ik + 1)\frac{\be}{2}\hbar \om_-}}\right) dk.
\eeq
 This  integral is given
as a series  by using the residue theorem.
One can easily see  that the numbers $(2m+1)i$, $m\in \mathbb Z$ are simple
pole of $\cosh{\pi\over2}k$,  and $i+{4\pi m\over(\beta\hbar\om_+)}$,
$i+{4\pi m\over(\beta\hbar\om_-)}$, $m\in \mathbb Z^{\ast}$
are simple or double poles of $\Theta(k)$. Now we can consider two
case, the first one where
\beq
\al\in\left]-\infty,\log\left({\mu\over\hbar\om}-\sq{\left({\mu\over\hbar\om}\right)^2-1}\right)\right]
\cup\left[\log\left({\mu\over\hbar\om}+\sq{\left({\mu\over\hbar\om}\right)^2-1}\right),+\infty\right[
\eeq
and here  we take an integration path lying in the lower half-plane and
involving only the simple poles $(2m+1)i$, $m< 0$.
It leads to the result
\beq
\Om(B,T,\al) = \frac{1}{4\be}\sum_{m=1}^{\infty} \frac{(-1)^m}{m}
\frac{e^{\beta \mu m}} {\sinh{(\frac{\be}{2}\hbar \om_+m)}\sinh{(\frac{\be}{2}\hbar \om_-m)}}
\label{4.20}
\eeq
%where the new frequencies are `` $\om_{\pm}=\om e^{\al}\pm \om_c$''.
In the second case where
\beq
\al\in\left[\log\left({\mu\over\hbar\om}-\sq{\left({\mu\over\hbar\om}\right)^2-1}\right),
\log\left({\mu\over\hbar\om}+\sq{\left({\mu\over\hbar\om}\right)^2-1}\right)\right]
\eeq
an integration path in the upper half-plane is chosen. It encircles all the other poles: $(2m+1)i$, $m\geq 0$,
$i+4\pi m/(\be\hbar\om_+)$, $i+4 \pi m/(\be\hbar\om_-)$, $m\in \mathbb
Z^{\ast}$.
We present the result in a manner which will render apparent the various regimes
\beq
\begin{array}{cccccccc}
\Om &=& &(\Om_L + \Om_{01})& + &\Om_{02}& + &\Om_{\mbox{\scriptsize osc}}\\
&=&2\pi i\Bigg(&\overbrace{a_{-1}(i)}& +&\overbrace{\sum_{m\geq 1}a_{-1}\left[(2m+1)i\right]}&
 +&\overbrace{\sum_{m_{\pm}\not= 0}
 \left[a_{-1}(i+\frac{4\pi}{\be\hbar\om_{\pm}}m_{\pm})\right]}\Bigg).
\end{array}
\eeq
where $\Om_{L}(B,\al)$ is given by
\beq
\Om_{L}={\mu\om_c^2\over24\lam} ={\mu\over24}
\left({\om_c^2\over\om^2-\om_c^2-2\om\om_c\sinh\al}\right)
\eeq
and $\Om_{01}(B,T,\al)$ reads as
\beq
\Om_{01}=-{\mu\over6}\left[\left({\mu\over\hbar\sq{\om^2-\om_c^2-2\om\om_c\sinh\al}}
\right)^2+\pi^2\left({k_BT\over\hbar\sq{\om^2-\om_c^2-2\om\om_c\sinh\al}}\right)^2-{1\over2}\right].
\eeq

Now let us consider an approximation such that
$\left({2\om\om_c\over\om^2-\om_c^2}\right)\sinh\al\ll1$. This allows us to
write the expansion
\beq\label{napprox}
{1\over\om_0^2-2\om\om_c\sinh\al}\simeq{1\over\om_0^2}+\left({2\om\om_c\over\om_0^4}\right)\sinh\al
\eeq
in the first order. Consequently, we find
This gives $\Om_{L}(B,\al)$ as
\beq
\Om_{L}(B,\al)={\mu\over24}\left({\om_c\over\om_0}\right)^2+
{\mu\over12}\left({\om_c\over\om_0^2}\right)^2\om\om_c\sinh\al. %=\Om_L+\Om(\al)
\eeq
This result can interpreted in different ways. Indeed, if we forget about the $\al$
appearing, we can reach the same conclusion as in~\cite{Gazeau}. Indeed,
the first term is at the origin of the Landau diamagnetism and gives
the susceptibility
%\beq
%\Om_L(B)={\mu\over24}\left({\om_c\over\om_0}\right)^2=-{1\over2}\chi_LH^2
%\eeq
%where
\beq
\chi_L=-{1\over3}\mu\left({\mu_B\over\hbar\om_0}\right)^2=-{1\over3}D_0\mu_B^2
\eeq
%is the Landau diamagnetic susceptibility.
where the coefficient
$D_0={\mu\over(\hbar\om_0)^2}$ can be interpreted as the density of
states at Fermi energy. Note that, the value of $\chi_L$ is equal to
one third of the one  $\chi_p$ found in (\ref{chi}).
On the other hand, $\al$ can be adjusted to get another contribution to
the susceptibility. More precisely, we can define $\al$ in terms of the
inverse of squared magnetic field to absorb the term $\om\om_c$ and therefore
get a a correction to the standard Landau diamagnetism. Otherwise, we can even
reproduce  $\chi_p$ simply by making an appropriate choice of $\al$. Indeed, fixing
$\sinh(\al) = \frac{\ga\om_0}{\om\om_c}$ to end up with (\ref{chi}),
where $\ga$ is constant that can be fixed easily. This show how
the obtained results generals and allow to deduce interesting properties.

According to (\ref{napprox}),
$\Om_{01}(\al)$ becomes
\beq
\Om_{01}(\al)=-{\mu\over6}\left[\left({\mu\over\hbar\om_0}\right)^2+
\pi^2\left({k_BT\over\hbar\om_0}\right)^2-{1\over2}\right]-
{\mu\over3}\left[\left({\mu\over\hbar\om_0^2}\right)^2+
\pi^2\left({k_BT\over\hbar\om_0^2}\right)^2\right]
\left(\om\om_c\over\om_0^4\right)\sinh\al.
\eeq
$\Om_{02}(\al)$ reads as
\beq
\Om_{02}(\al) = \frac{1}{4\be}\sum_{m=1}^{\infty} \frac{(-1)^m}{m}
\frac{e^{-\beta \mu m}}
     {\sinh{\left[\frac{\hbar\left(\om e^{\al}+\om_c\right)}
{2k_BT}m\right]}\sinh{\left[\frac{\hbar\left(\om e^{-\al}-
\om_c\right)}{2k_BT}m\right]}}.
\label{4.21}
\eeq
Note  that, when $\al=0$ we get $\lambda=\om_0$, which leads to  recover the
result obtained in~\cite{Gazeau}. It becomes negligible at low temperature regime $k_BT\ll \mu$.
The sum of $\Om_L$ and $\Om_{01}$ is analogue to the term $A$ in (\ref{4.11})
and $\Om_{02}$ corresponds to $S_{\pm}$.
The last term is responsible for the oscillatory behavior. If
$\om_+/\om_-$  is irrational values,  we have
\begin{eqnarray}\lb{4.26}
\nonumber \Om_{\mbox{\scriptsize osc}}(\al)&=&
{1\over2\be} \sum_{m=1}^{\infty} {(-1)^m\over m}
\left\lbrack{\sin{\left[{2 \mu\over\hbar( \om e^{-\al}-\om_c)}\pi m\right]}\over
\sin{\left[{\om e^{\al}+\om_c\over \om e^{-\al}-\om_c}\pi m\right]}\,
\sinh{\left[{2k_B T\over\hbar( \om e^{-\al}-\om_c)}\pi^2 m\right]}} \right.\\
 &+& \left. {\sin{\left[{ 2\mu\over\hbar ( \om e^{\al}+\om_c)}\pi m\right]
}\over\sin{\left[{\om e^{-\al}-\om_c\over\om e^{\al}+\om_c}\pi m\right]}\,
\sinh{\left[{2 k_B T\over\hbar ( \om e^{\al}+\om_c)}\pi^2 m\right]}} \right\rbrack
\equiv \Om_{\mbox{\scriptsize osc}}^-(\al) + \Om_{\mbox{\scriptsize osc}}^+(\al).
\end{eqnarray}

%%%%%%%%%%%%%%%%%%%%%%%%%%%%%%%%%%%%%%%%%%%%%%%%%%%%%%%%%%%%%%%%%%%%%%%%%%%%%%%%%
\subsection{Average number of electrons}
%%%%%%%%%%%%%%%%%%%%%%%%%%%%%%%%%%%%%%%%%%%%%%%%%%%%%%%%%%%%%%%%%

In this section, we will exploit the formula's (\ref{4.20})-(\ref{4.26}) to
obtain the exact expressions of the average number of electrons and
the magnetic moment. We will restrict ourselves to the more realistic case:
$\mu \leq \hbar\om/2 $.
The average number of electrons is easily derived by taking the derivative of
$-\Om$ with respect to $\mu$. It is found to be
\beq
\langle N_e\rangle=-\partial_{\mu}\Om_L(\al)-\partial_{\mu}\Om_{01}(\al)-\pa_{\mu}\Om_{\mbox{\scriptsize
 osc}}^-(\al)-\partial_{\mu}\Om_{\mbox{\scriptsize
  osc}}^+(\al).
\eeq
With the straightforward calculation, we find
\beq\label{nnple}
\langle N_e(\al)\rangle = \langle N_e\, \rangle_{L} + \langle N_e\,\rangle_{01}
+\langle N_e\, \rangle_{02} + \langle N_e\,\rangle_{\mbox{\scriptsize osc}}^-
+\langle N_e\, \rangle_{\mbox{\scriptsize osc}}^+
\eeq
where different portions are give by
\begin{eqnarray}%\label{nnple}
\langle N_e\, \rangle_{L} &=&
-{1\over24}\left({\om_c\over\om_0}\right)^2+
{1\over2}
\left[\left({\mu\over\hbar\om_0}\right)^2+{\pi^2\over3}
\left({k_BT\over\hbar\om_0}\right)^2-{1\over6}\right]\\
\langle N_e\, \rangle_{01} & =&
\left[\left({\mu\over\hbar\om_0^2}\right)^2+{\pi^2\over3}
\left({k_BT\over\hbar\om_0^2}\right)^2-{\om_c^2\over12}\right]{\om\om_c\over\om_0^4}\sinh\al\\
\langle N_e\, \rangle_{02} & =&
{1\over4}\sum_{m=1}^{\infty} (-1)^m
{e^{-\beta \mu m}
     \over\sinh{\left[{\hbar\left(\om
     e^{\al}+\om_c\right)\over2k_BT}m\right]}\,\sinh{\left[{\hbar\left(\om
     e^{-\al}-\om_c\right)\over2k_BT}m\right]}}\\
\langle N_e\,\rangle_{\mbox{\scriptsize osc}}^- &=& -
\pi \sum_{m=1}^{\infty}(-1)^m
\left.{k_BT\over\hbar\left(\om e^{-\al}-\om_c\right)}
{\cos{\left[{2 \mu\over\hbar( \om e^{-\al}-\om_c)}\pi m\right]}\over
\sin{\left[{\om e^{\al}+\om_c\over \om e^{-\al}-\om_c}\pi m\right]}\,
\sinh{\left[{2k_B T\over\hbar( \om e^{-\al}-\om_c)}\pi^2 m\right]}} \right.\\
\langle N_e\,\rangle_{\mbox{\scriptsize osc}}^+ &=& -
\pi \sum_{m=1}^{\infty}(-1)^m
 \left.{k_BT\over\hbar\left(\om e^{\al}+\om_c\right)} {\cos{\left[{ 2\mu\over\hbar
( \om e^{\al}+\om_c)}\pi m\right]}\over\sin{\left[{\om e^{-\al}-\om_c\over\om e^{\al}+\om_c}\pi m\right]}\,
\sinh{\left[{2 k_B T\over\hbar ( \om e^{\al}+\om_c)}\pi^2 m\right]}}\right..
\end{eqnarray}

The above results are general in sense that the standard solutions can be
recovered. Indeed, requiring that $\al=0$, we show
%The average number of electrons is easily derived by taking the derivative of
%$-\Om$ with respect to $\mu$. It is found to be:
\begin{eqnarray}
\nonumber \langle N_e\,\rangle &=&
-\frac{1}{24}\left(\frac{\om_c}{\om_0}\right)^2
+\frac{1}{2}\left\lbrack\left(\frac{\mu}{\hbar \om_0}\right)^2
+\frac{\pi^2}{3}\left(\frac{k_BT}{\hbar\om_0}\right)^2
-\frac{1}{6}\right\rbrack \\
\nonumber & & +\frac{1}{4} \sum_{m=1}^{\infty} (-1)^m
\frac{e^{-\be \mu m}}{\sinh{(\frac{\be}{2}\hbar\om_+m)}\,
\sinh{(\frac{\be}{2}\hbar \om_-m)}} \\
\nonumber & &- \pi\sum_{m=1}^{\infty} (-1)^m
\left\lbrack \frac{k_B T}{\hbar\om_-}
\frac{\cos{(\frac{2 \mu}{\hbar \om_-}\pi m)}}{\sin{(\frac{\om_+}{\om_-}\pi m)}\,
\sinh{(\frac{2k_B T}{\hbar\om_-}\pi^2 m)}}
+\frac{k_B T}{\hbar \om_+ }\frac{\cos{(\frac{2\mu}{\hbar \om_+}\pi m)}}
{\sin{(\frac{\om_-}{\om_+}\pi m)}\,
\sinh{(\frac{2 k_B T}{\hbar \om_+}\pi^2 m)}} \right\rbrack. % \\
%&\equiv& \langle N_e\, \rangle_{L} + \langle N_e\,\rangle_{01}
%+\langle N_e\, \rangle_{02} + \langle N_e\,\rangle_{\mbox{\scriptsize osc}}^-
%+\langle N_e\, \rangle_{\mbox{\scriptsize osc}}^+. \label{5.1}
\end{eqnarray}

%%%%%%%%%%%%%%%%%%%%%%%%%%%%%%%
\subsection{Magnetic moment}
%%%%%%%%%%%%%%%%%%%%%%%%%%%%%%
The magnetic moment is found by the conversely of the derivative of the thermodynamical
potential by the magnetic field. Then,  we have
\begin{equation}
M(B,T,\al) =
\nonumber
-2\mu_B\left({\pa\Om_L(\al)\over\pa\hbar\om_c}\right)-2\mu_B
\left({\pa\Om_{01}(\al)\over\pa\hbar\om_c}\right)-2\mu_B
\left({\pa\Om_{02}(\al)\over\pa\hbar\om_c}\right)-2\mu_B\left({\pa\Om_{\mbox{\scriptsize
      osc}}(\al)\over\pa\hbar\om_c}\right).
\end{equation}
It is equivalent to
\begin{equation}
M(B,T,\al) =
2\mu_B\left[M_L(\al)+M_{01}(\al)+M_{02}(\al)+M_{\mbox{\scriptsize
      osc}}^-(\al)+M_{\mbox{\scriptsize
      osc}}(\al)^+\right]
\end{equation}
where different quantities read as
% $M_L(B,T,\al)$ is
\beqar
M_L(B,T,\al) &=& -{\mu\over12\hbar\om_0}\left({\om_c\over\om_0}\right)-
{\mu\over12\hbar\om_0^4}\left(3\om+{\om_c^2\over\om}\right)\om_c^2\sinh\al \\
M_{01}(B,T,\al)& =& {\mu\over3\hbar\om}\left[\left({\mu\over\hbar\om_0^2}\right)^2+
\pi^2\left({k_BT\over\hbar\om_0^2}\right)^2\right]\left({\om^2+\om_c^2}\right)\sinh\al.
\\
M_{02}(B,T,\al) &=& {1\over8}\sum_{m=1}^{\infty}(-1)^m e^{-\beta \mu
  m}\left[{\coth\left({\hbar\om_-\over2k_BT}m\right)-
\coth\left({\hbar\om_+\over2k_BT}m\right)\over\sinh
\left({\hbar\om_+\over2k_BT}m\right)\sinh\left({\hbar\om_-\over2k_BT}m\right)}\right].
\\
M_{\mbox{\scriptsize osc}}^+(B,T,\al) &=&
{k_BT\over\hbar\om_+}\sum_{m=1}^{\infty}{(-1)^m\sin\left({2
    \mu\over\hbar\om_+}\pi m\right)\over\sin\left({\om_-\over\om_+}\pi
  m\right)\sin\left({2k_BT\over\hbar\om_+}\pi^2 m\right)}\times\Bigg[{\mu\pi\over\hbar\om_+}\cot\left({2\mu\over\hbar\om_+}\pi
  m\right)\nonumber\\
&&
-{\om\cosh\al\over\om_+}\pi\cot\left({\om_-\over\om_+}\pi
  m\right)-{k_BT\over\hbar\om_+}\pi\coth\left({2k_BT\over\hbar\om_+}\pi^2m\right)\Bigg]
\\
M_{\mbox{\scriptsize osc}}^-(B,T,\al) &=&
-{k_BT\over\hbar\om_-}\sum_{m=1}^{\infty}{(-1)^m\sin\left({2
    \mu\over\hbar\om_-}\pi m\right)\over\sin\left({\om_+\over\om_-}\pi
  m\right)\sin\left({2k_BT\over\hbar\om_-}\pi^2 m\right)}\times\Bigg[{\mu\pi\over\hbar\om_-}\cot\left({2\mu\over\hbar\om_-}\pi
  m\right)\nonumber\\
&&
-{\om\cosh\al\over\om_-}\pi\cot\left({\om_+\over\om_-}\pi
  m\right)-{k_BT\over\hbar\om_-}\pi\coth\left({2k_BT\over\hbar\om_-}\pi^2m\right)\Bigg].
\eeqar

In the case where $\al=0$, we end up with
%The magnetic moment is decomposed into four parts and is expressed in Bohr
%magneton units:
%\begin{eqnarray}
%\nonumber M &=& \chi_L H
%-2\mu_B\left(\frac{\pa\Om_{02}}{\pa\hbar\om_c}\right)_{\mu}
%-2\mu_B\left(\frac{\pa \Om_{\mbox{\scriptsize osc}}}{\pa\hbar\om_c}\right)_{\mu}
%\\
%&\equiv & 2\mu_B({\cal M}_L +{\cal M}_0+{\cal M}_{\mbox{\scriptsize osc}}^- +
 %{\cal M}_{\mbox{\scriptsize osc}}^+),
%\end{eqnarray}
%where
\begin{eqnarray}
{\cal M}_L &=& \frac{-\mu}{12\hbar\om_0}\left( \frac{\om_c}{\om_0} \right)
        \equiv \frac{1}{2\mu_B} \chi_L H ,\\
{\cal M}_0 &=& \frac{1}{8\om}\sum_{m=1}^{\infty}
          (-1)^m e^{-\beta \mu m } \frac{
                 [ \om_+\coth{(\beta\hbar\om_+m/2)}
                  -\om_-\coth{(\beta\hbar\om_-m/2)}]}
                {\sinh{(\beta\hbar\om_+m/2)}\sinh{(\beta\hbar\om_-m/2)}},
\end{eqnarray}
and, for the irrational case $\om_+/\om_- \not\in \mathbb{Q}$,
\begin{eqnarray}
{\cal M}_{\mbox{\scriptsize osc}}^- &=&
-\frac{k_BT}{\hbar\om} \sum_{m=1}^{\infty}
 \frac{(-1)^m \sin{(2\pi m\mu/(\hbar\om_-))}}
                  {\sin{(\pi m\om_+/\om_-)} \sinh{(2\pi^2mk_BT/(\hbar\om_-))}}
             \times \\ \nonumber
          & &\left[
 \frac{\pi\mu}{\hbar\om_-}\cot{\left(2\pi m\frac{\mu}{\hbar\om_-} \right)}
-\frac{\pi\om_+}{\om_-}\cot{\left(\pi m\frac{\om_+}{\om_-} \right)}
-\frac{\pi^2k_BT}{\hbar\om_-}\coth{\left(2\pi^2m\frac{k_BT}{\hbar\om_-}\right)}
             \right],\\
%\nonumber
{\cal M}_{\mbox{\scriptsize osc}}^+ &=&
\frac{k_BT}{\hbar\om} \sum_{m=1}^{\infty}
              \frac{(-1)^m \sin{(2\pi m\mu/(\hbar\om_+))}}
                  {\sin{(\pi m\om_-/\om_+)}\sinh{(2\pi^2mk_BT/(\hbar\om_+))}}
              \times \\ \nonumber
          & &\left[
 \frac{\pi\mu}{\hbar\om_+}\cot{\left(2\pi m\frac{\mu}{\hbar\om_+} \right)}
-\frac{\pi\om_-}{\om_+}\cot{\left(\pi m\frac{\om_-}{\om_+} \right)}
-\frac{\pi^2k_BT}{\hbar\om_+}\coth{\left(2\pi^2m\frac{k_BT}{\hbar\om_+}\right)}
             \right].
\end{eqnarray}
We will not give the expressions of ${\cal M}^{\pm}_{\mbox{\scriptsize osc}}$
in the rational case because the magnetization is a continuous function of
$\om_c$ and its behavior can be fully understood from the irrational one.

In the end note that,
the temperature scale is compared to the two natural modes $\om_{\pm}$ of the
system and draws three possible intrinsic regimes: high temperature regime
$k_B T > \hbar \om_+$, low temperature regime $k_B T < \hbar \om_-$, and
intermediate temperature regime $\hbar \om_- < k_B T < \hbar \om_+$.
Remember that we work in the large electron number region: $\mu > \hbar\om/2$.

%%%%%%%%%%%%%%%%%%%%%%%%%%%%%%%%%%%%%%
\section{Conclusion}
%%%%%%%%%%%%%%%%%%%%%%%%%%%%%%%%%%

We started by formulating our problem in two-dimensional
space where two coupled harmonic
oscillators living on. Subsequently, we introduced a minimal coupling
to generate another interacting system that is
studied. After rescaling different variables, we showed that
it is possible to get a
diagonalized Hamiltonian. In fact, this is done
by making use
%The energy solution is obtained after making use
of an unitary transformation. It was helpful in sense
that the eigenvalues and their wavefunctions are obtained
in simple way in terms of the coupling parameter $\al$.

The fact that the energy spectrum solutions are $(\al,B)$
dependent, we discussed their underlying properties.
More precisely, four limiting cases have been
investigated, which are week and strong parameters
$(\al,B)$. In particular, we noticed that by fixing $\al$,
some model can be recovered, these concern for instance the
Landau Hamiltonian in two dimensions and harmonic oscillator
in one-dimension. These allowed us to conclude that
by adjusting the coupling parameter, one can derive
other interesting
other solutions.

%We analyzed the basic feature of two coupled harmonic
%oscillators in the presence of an external magnetic field.
%After making use of an unitary transformation, we ended up
%with a solvable Hamiltonian of the system. This was helpful in sense
%that the eigenvalues and their wavefunctions are obtained
%in simple way. Taking into account of the week and
%strong magnetic limits, we derived different properties
%of the energy spectrum and intermediate is also discussed.

To investigate different issues related to the considered
system, we constructed the corresponding coherent states,
which are obtained to be coupling parameter dependent.
These are used to evaluate the thermodynamical potential
by adopting two different methods. First method employed the
Berezin--Lieb inequalities to obtain an approximate form.
Using this to determine the average number of electrons $\langle N_e\rangle$ and
the magnetization as well as underline their properties
in terms of the the limiting cases $(\al\ll1, \al\lga\infty)$
as well as week and strong magnetic field limits.
In fact, by treating the limit $\al\ll 1$ a correction to
 $\langle N_e\rangle$ is obtained.
More importantly, we showed that $\al$ can be tuned to reproduce both
the orbital paramagnetism and the Landau diamagnetism in such limit.
In fact, we derived a general magnetic moment that can be
fixed to reproduce different results and end up with
some conclusions

In the Second method, we employed some mathematical toy to determine the
exact formula of the thermodynamical potential and therefore evaluated
different physical quantities. More precisely, the Fermi--Dirac trace formulas
is used and the average number of electrons as well
as the magnetic moment are calculated.
After evaluating the susceptibility, we found that there is a correction to the Landau
diamagnetism, which is $\al$-dependent. Again by fixing
the parameter, other results can be obtained and in particular
for $\al=0$  standard results~\cite{Gazeau} is easily
recovered.

Some interesting questions remain to be solved for the present system.
In fact, first concerns the temperature limits of the thermodynamical potential
obtained in terms of the second method. This can also be investigated further by
considering all limiting cases of the couple $(\al,B)$. Second is related to
discuss the spatial density of current.
Finally, a numerical study if the obtained results is much needed to
give another comparisons with already published results.

%%%%%%%%%%%%%%%%%%%%%%%%%%%%%%%%%%%%%
\section*{Acknowledgment}
%%%%%%%%%%%%%%%%%%%%%%%%%%%%%%%%%%%%%

This work was completed during AJ visit to
Max Planck Institute for the Physics of Complex Systems, Dresden. He would
like to thank the Institute for %the financial support and
the warm hospitality. He is also thankful to Mr. M. Said for his
administrative help.

%%%%%%%%%%%%%%%%%%%%%%%%%%%%%%%%%%%%%%%%%%%%%%%%%%%%%%%%%%%%%
\section*{Appendix: Fermi-Dirac trace formulas}
%%%%%%%%%%%%%%%%%%%%%%%%%%%%%%%%%%%%%%%%%%%%%%%%%%%%%%%%%%%%%%

It is well known that, like the Gaussian function, the function
$\mbox{sech}{x} = 1/\cosh{x}$ is a fixed point for the Fourier transform in
the Schwartz space:
\beq
\frac{1}{\cosh{\sqrt{\frac{\pi}{2}}x}}
= \frac{1}{\sqrt{2\pi}} \int_{-\infty}^{+\infty}
\frac{e^{-ixy}}{\cosh{\sqrt{\frac{\pi}{2}}y}}\, dy.
\eeq
Hence, given an Hamiltonian ${\cal  H}$, we can write for the corresponding Fermi operator:
\beq
f({\cal H})\equiv\frac{1}{1+e^{\be({\cal H}-\mu)}}=
\int_{-\infty}^{+\infty}\frac{e^{-(ik+1) \frac{\be}{2} ({\cal H}-\mu)}}
{4\cosh{\frac{\pi}{2}k}}\, dk. \label{a.2}
\eeq
Similarly, we can write for the thermodynamical potential operator:
\beq
-\frac{1}{\be}\log{(1+e^{-\be({\cal H} - \mu)})} = -\frac{1}{\be}\int_{-\infty}^{+\infty} \frac{e^{-(ik +
1)\frac{\be}{2}({\cal  H} - \mu)}}{(2\cosh{\frac{\pi}{2}k})(ik+1)}\, dk.
\eeq
Therefore, the average number of fermions and the thermodynamical potential can be written (at least formally) as follows:
\beqar
\langle N \rangle &=& \mbox{Tr}f({\cal  H})
=\int_{-\infty}^{+\infty}
\frac{e^{(ik+1)\frac{\be\mu}{2}}}{4\cosh{\frac{\pi}{2}k}}\Theta(k)\,dk
\label{a.4} \\
\Om &=& \mbox{Tr}(-\frac{1}{\be}\log{(1+ e^{-\be({\cal  H} - \mu)})})
=-\frac{1}{\be}\int_{-\infty}^{+\infty}
\frac{e^{(ik+1)\frac{\be\mu}{2}}}{(2\cosh{\frac{\pi}{2}k})(ik+1)}\Theta(k)\,dk
\label{a.5}
\eeqar
where $\Theta$ designates the function
\beq
\Theta (k) =\mbox{Tr} (e^{-(ik +
1)\frac{\be}{2}{\cal  H}}).
\label{a.6}
\eeq
Observe that  $(2m+1)i, \ m\in \mathbb{Z}$ are (simple) poles for the function
$1/\cosh{\frac{\pi}{2}k}$ and $i$ is a pole for the functions $\Theta(k)$
and $1/(ik +1)$.
These Fourier integrals can be evaluated by using residue theorems
if the integrand functions
$\Phi_1(k)=\Theta (k)/\cosh{\frac{\pi}{2}k}$ and
$\Phi_2(k)=\Theta (k)/((ik+1)\cosh{\frac{\pi}{2}k})$
satisfy the Jordan Lemma, that is,
$\Phi_1(R e^{i\te}) \leq g(R)$, $\Phi_2(R e^{i\te})\leq h(R)$, for all
$\te \in \lbrack 0, \pi \rbrack$, and
$g(R)$ and $h(R)$ vanish as $R\to \infty$. The quantities
$\langle N \rangle$ and $\Om$ are then formally given by
\beq
2\pi i \left\lbrack a_{-1}(i)+\sum_{m=1}^{\infty}a_{-1}((2m+1)i)
+\sum_{\nu} a_{-1}(k_{\nu})  \right\rbrack
\label{a.7}
\eeq
where $a_{-1}(\cdot)$ denotes the residue of the involved integrand at pole
$(\cdot)$, and the $k_{\nu}$'s are the poles (with the exclusion of the pole $i$) of $\Theta(k)$ in the
complex $k$-plane.

We now introduce the spectral resolution of the (bounded below) self-adjoint
operator ${\cal  H}$:
\beq
\varphi({\cal H})=\int_{-\infty}^{+\infty} \varphi(\lambda)\, E(d\lambda)
\eeq
where $\varphi$ is a complex-valued function and
$E_{\lambda} = \int_{-\infty}^{\lambda}\, E(d\lambda)$
is the resolution of the identity for the Hamiltonian ${\cal  H}$.
Define the density of states  $\nu(\lambda)$ as
$\mbox{Tr}E(d\lambda)/d\lambda $.
The trace formula ensues:
\beq
\mbox{Tr}\varphi({\cal H})=
\int_{-\infty}^{+\infty}\varphi(\lambda)\,\nu(\lambda)\, d\lambda. \label{a.9}
\eeq
Let us now introduce the weighted density of states
$ w(\lambda) = e^{-\frac{\be}{2} \lambda} \nu (\lambda)$
and its Fourier transform
\beq
\hat{w} (k) = \frac{1}{\sqrt{2 \pi}}\int_{-\infty}^{+\infty} e^{-ik \lambda} w(\lambda)\, d\lambda.
\eeq
Then, from (\ref{a.4}), (\ref{a.5}) and (\ref{a.9}), we can represent
$\langle N \rangle$ and $\Om$ as follows:
\beqar
\langle N \rangle &=&\sqrt{2\pi} \int_{-\infty}^{+\infty}
\frac{e^{(ik + 1)\frac{\be \mu}{2}}}{4\cosh{\frac{\pi}{2}k}}
\hat{w}(\frac{\be }{2} k)\, dk
=\frac{\pi}{\be} e^{\frac{\be \mu}{2}}\widehat{{\cal Z}}_1 (- \mu) \\
\Om &=& -\frac{\sqrt{2\pi}}{\be} \int_{-\infty}^{+\infty}
\frac{e^{(ik + 1)\frac{\be\mu}{2}}}{(2\cosh{\frac{\pi}{2}k})(ik+1)}
\hat{w}(\frac{\be }{2} k)\, dk
= -\frac{2\pi}{\be^2} e^{\frac{\be \mu}{2}} \widehat{{\cal Z}}_2 (- \mu)
\eeqar
where we have introduced the weighted functions
\beq
{\cal Z}_1=\mbox{sech}(\frac{\pi}{\be} k) \hat{w} (k), \qquad
{\cal Z}_2 = \mbox{sech}(\pi k/\be) (i2k/\be+1)^{-1} \hat{w}(k).
\eeq


\begin{thebibliography}{1}

\bibitem{landau} L.D. Landau
{\it Z. Phys.} {\bf 64} (1930) 629.

\bibitem{ber} K. Richter, D. Ullmo and R.A. Jalabert, {\it Phys. Rep.} {\bf 276} (1996) 1.
\bibitem{fks} M. Combescure and D. Robert, {\it Rev. Math. Phys.} {\bf 13} (2001) 1055.

\bibitem{Plischke} M. Plischke and B. Bergersen, Equilibrium Statistical Physics, Third Edition,
 (World Scientific, Singapore, 2006).

\bibitem{yfu} D. Yoshioka and H. Fukuyama,
{\it J. Phys. Soc. Jap.} {\bf 61} (1992) 2368.

%\bibitem{ber}  F.A. Berezin,
%{\it Izv. Akad.  SSSR Ser. Mat.} {\bf 6} (1972) 1134;
%{\it Commun. Math.  Phys.} {\bf 40} (1975) 153.

%\bibitem{fks}  D.H. Feng, J.R. Klauder and  M. Strayer (eds):
%{\em Coherent States: Past, Present and Future} ({\em Proc. Oak Ridge 1993}),
%(World Scientific, Singapore, 1994).

\bibitem{ifu} Y. Ishikawa and H. Fukuyama,
{\it J. Phys. Soc. Jap.} {\bf 68} (1999) 2405.



\bibitem{Gazeau}   J.P. Gazeau, P.Y. Hsiao and A. Jellal, {\it Phys. Rev.} {\bf B65} (2002) 094427,
 {\sf cond-mat/0101338}.





\bibitem{kim}
Y.S. Kim, M.E. Noz and S.H. Oh,
%{\it A Simple Method for Illustrating the Difference between
%the Homogeneous and Inhomogeneous Lorentz Groups},
{\it Am. J. Phys.} {\bf 47} (1979) 892.

\bibitem{kim2}
Y.S. Kim and M.E. Noz, ``{Theory and Applications of the
Poincar\'e Group}'', (Reidel, Dordrecht, 1986).

\bibitem{kim3}D. Han, Y.S. Kim and M.E. Noz,
%{\it Lorentz-Squeezed Hadrons
%and Hadronic Temperature},
{\it Phys. Lett.}  {\bf A144} (1989) 111.

\bibitem{kim4}Y.S. Kim,
%{\it Observable Gauge Transformations in the Parton Picture},
{\it Phys. Rev. Lett. } {\bf 63}, 348 (1989).

\bibitem{kim5}Y.S. Kim and E.P. Wigner,
%{\it Entropy and Lorentz Transformations},
{\it Phys. Lett. } {\bf A147} (1990) 343.

\bibitem{kim6}D. Han, Y.S. Kim and M.E. Noz,
%{\it Linear Canonical Transformations of Coherent and
%Squeezed States in the Wigner Phase Space III.  Two-mode States},
{\it Phys. Rev. } {\bf A41}  (1990) 6233.

\bibitem{kim7}Y.S. Kim and M.E. Noz, "{ Phase Space Picture of Quantum
Mechanics}", (World Scientific, Singapore, 1991).

\bibitem{kimg}D. Han, Y.S. Kim and M.E. Noz,
%{\it O(3,3)-like Symmetries of
%Coupled Harmonic Oscillators}
{\it J. Math. Phys.} {\bf 36}  (1995) 3940; ibid
{\it Am. J. Phys.} {\bf 67} (1999) 61.

%\bibitem{ishikawa} Y. Ishikawa and H. Fukuyama, {\it J. Phys> Soc. Jap.}
%{\bf 68} (1999) 2405.

\bibitem{jellal}  A. Jellal, E.H. El Kinani and M. Schreiber,
{\it  Int. J. Mod. Phys.} {\bf A20} (2005) 1515,  {\sf hep-th/0309105}.


\bibitem{jellal2} A. Jellal, {\it Nucl. Phys.} {\bf B804} (2008) 361, {\sf arXiv:0709.4126}.

\bibitem{prange} For instance see R.E. Prange and S.M. Girvin (editors), ”The Quantum Hall Effect”
(Springer, New York 1990).

\end{thebibliography}
\end{document}